\documentclass[showpacs,aps,preprint,amsmath,amssymb,superscriptaddress]{revtex4}
\usepackage{graphicx}
\usepackage{dcolumn}
\begin{document}
\title{Quasiparticle self-energy and many-body effective mass enhancement 
in a two-dimensional electron liquid}
\author{R. Asgari}
\affiliation{NEST-INFM and Classe di Scienze, Scuola Normale Superiore, I-56126 Pisa, Italy}
\affiliation{Institute for Studies in Theoretical Physics and Mathematics, Tehran 19395-5531, Iran}
\author{B. Davoudi\footnote{Present address: D\'epartement de Physique and Centre de Recherche en Physique du Solide, 
Universit\'e de Sherbrooke, Sherbrooke, Qu\'ebec, Canada J1K 2R1}}
\affiliation{NEST-INFM and Classe di Scienze, Scuola Normale Superiore, I-56126 Pisa, Italy}
\affiliation{Institute for Studies in Theoretical Physics and Mathematics, Tehran 19395-5531, Iran}
\author{M. Polini\footnote{e-mail: m.polini@sns.it}}
\affiliation{NEST-INFM and Classe di Scienze, Scuola Normale Superiore, I-56126 Pisa, Italy}
\author{G.F. Giuliani}
\affiliation{Physics Department, Purdue University, West Lafayette, Indiana 47907, USA}
\author{M.P. Tosi}
\affiliation{NEST-INFM and Classe di Scienze, Scuola Normale Superiore, I-56126 Pisa, Italy}
\author{G. Vignale}
\affiliation{Department of Physics and Astronomy, University of Missouri-Columbia, Columbia, Missouri 65211, USA}
\begin{abstract}
Motivated by a number of recent experimental studies 
we have revisited the problem of the microscopic calculation 
of the quasiparticle self-energy and many-body effective mass 
enhancement in a two-dimensional electron liquid. Our systematic study is based on 
the many-body local fields theory and takes advantage of the results of 
the most recent Diffusion Monte Carlo calculations of the static charge and spin 
response of the electron liquid. We report extensive calculations
of both the real and imaginary parts of the 
quasiparticle self-energy. We also present results for the many-body effective mass 
enhancement and the renormalization constant over an extensive range of electron density. 
In this respect we critically examine the relative merits 
of the on-shell approximation versus the self-consistent solution of the Dyson equation. 
We show that in the strongly-correlated regime 
a solution of the Dyson equation proves necessary in order to obtain a 
well behaved effective mass. The inclusion of both charge- and spin-density fluctuations 
beyond the Random Phase Approximation is indeed crucial 
to get reasonable agreement with recent measurements.
\end{abstract}
\pacs{71.10.Ca}
\maketitle

\section{Introduction}
\label{introduction}
An interacting electron gas on a uniform neutralizing background (EG)
is used as the reference system in most realistic calculations of
electronic structure in condensed-matter physics~\cite{ceperley_1999}.
At zero temperature there are only two relevant parameters for a disorder-free EG 
in the absence of quantizing magnetic fields and spin-orbital coupling: (i) the usual Wigner-Seitz density parameter $r_s=(\pi n_{\rm \scriptscriptstyle 2D} a^2_B)^{-1/2}$, $a_B=\hbar^2{\bar \kappa}/(m e^2)$ being the Bohr radius in the medium of interest with ${\bar \kappa}$ and $m$ appropriate dielectric constant and bare band mass respectively; and (ii) the degree of spin polarization $\zeta=|n_\uparrow-n_\downarrow|/n_{\rm \scriptscriptstyle 2D}$. Here 
$n_\sigma$ is the average density of particles with spin $\sigma=\uparrow,\downarrow$ and $n_{\rm \scriptscriptstyle 2D}=n_\uparrow+n_\downarrow$ is the total average density. 

Understanding the many-body aspects of this model
has attracted continued interest for many decades~\cite{singwi_1981,march_book}. 
The EG, unlike systems of classical particles, behaves like an ideal paramagnetic gas 
at high density ($r_s\ll 1$) and like a solid at low density~\cite{wigner} ($r_s\gg 1$). In the intermediate
density regime, which is relevant in three dimensions ($3D$) to conduction electrons in simple metals and in two dimensions ($2D$) to electrons in an inversion layer of a Si metal-oxide-semiconductor field-effect transistor (MOSFET) or in an AlGaAs/GaAs quantum well, perturbative techniques are not effective owing to the lack of a small expansion parameter. One has to take recourse to approximate semi-analytical methods, a number of which have been reviewed in Ref.~\onlinecite{singwi_1981}, or to Quantum Monte Carlo (QMC) simulation methods~\cite{ceperley_1980,tanatar,moroni_1992,kwon_1994,moroni_1995,
rapisarda,ortiz,senatore_1999,varsano,zong,attaccalite}.

Among the methods designed to deal with the intermediate density regime, of particular interest for its physical appeal and elegance is Landau's phenomenological theory~\cite{landau} dealing with low-lying excitations in a Fermi-liquid. Landau called such single-particle excitations quasiparticles (QP's) and postulated a one-to-one correspondence between them and the excited states of a non-interacting Fermi gas. He wrote the excitation energy of the Fermi-liquid in terms of the energies of the QP's and of their effective interactions. The QP-QP interaction function can in turn be used to obtain various physical properties of the system and can be parametrized in terms of experimentally measurable data. 

Quinn and Ferrell~\cite{quinn} provided a framework for the microscopic evaluation of the QP-QP interactions in the EG by means of the Random Phase Approximation (RPA). Next, Rice~\cite{rice} incorporated the vertex corrections in the RPA form of the electron self-energy by including the Hubbard~\cite{hubbard} many-body local-field. Some problems in Rice's theory were subsequently resolved by Ting, Lee, and Quinn~\cite{ting} in their theory of the quasi-$2D$ EG. All these approaches considered only the charge-density fluctuations while neglecting the effect of the spin-density fluctuations. 

A more detailed analysis that accounts for the vertex corrections associated with both types of fluctuations was carried out for an unpolarized EG in Refs.~\onlinecite{vignale_1985,zhu_1986,yarlagadda_1989}, where Kukkonen-Overhauser-like~\cite{kukkonen} effective interactions were obtained by different approaches. In particular, Yarlagadda and Giuliani~\cite{yarlagadda_1989} adopted a physically transparent approach termed renormalized Hamiltonian approach (RHA),  which will be extensively discussed in this paper. 
A few electrons from the EG are selected and called  ``test electrons", while the remaining EG is treated as a dielectric screening medium. As the test electrons move through this medium, they produce fluctuations in the density of spin-up and spin-down electrons, which provide virtual clothing and also screen their interactions. Thus, the dielectric mimics the true physical processes in an average way. Of course, the test electrons and the electrons of the medium are physically indistinguishable, and this must be taken into account when exchange effects are considered. At this point, after averaging over the coordinates of the screening medium, an effective renormalized Hamiltonian containing only the degrees of freedom of the clothed test electrons (or QP's) can be derived, under the assumption that the coupling with the medium occurs only {\it via} its charge- and spin-density fluctuations. The basic idea underlying the RHA was earlier developed in a beautiful paper 
by Hamann and Overhauser~\cite{hamann} within the RPA. Calculations based on these theories have been carried out for both $3D$~\cite{zhu_1986,rahman_1984} and $2D$~\cite{santoro_1989} systems. 

In a parallel theoretical development Ng and Singwi~\cite{ng}, starting from a Ward identity and performing a local approximation on the irreducible particle-hole interaction,  obtained an expression for the self-energy  in terms of the many-body local field factors associated with charge- and spin-density fluctuations. Equivalent results were later obtained by Yarlagadda and Giuliani~\cite{yarlagadda_1994,yarlagadda_gstar} by means of the RHA. These authors also took into account an infinitesimal degree of spin polarization which allowed them to carry out  calculations of the Landau Fermi-liquid parameters. 

The previous theoretical work suffers from two major shortcomings. (i) All available earlier calculations have adopted a static and oversimplified Hubbard-like model for the local-field factors, which do not have the appropriate behavior at both intemediate and large wave number $q$. We have corrected for these discrepancies (while still neglecting the frequency dependence of the local fields)  using recent parametrizations~\cite{davoudi_2001} for the static local fields of a $2D$ EG. (ii) The RHA briefly introduced above is an ``on-shell" theory (see Sect.~\ref{rha}) and, as we will show in this work, predicts a divergence of the effective mass with decreasing electron density. In the spirit of the work of Santoro-Giuliani~\cite{santoro_1989} and of Ng-Singwi~\cite{ng} we have kept in this work the full frequency dependence of the self-energy and carried out a self-consistent solution of the Dyson equation to find the proper QP excitation energy and QP properties. Comparing with Ref.~\onlinecite{santoro_1989}, we have released the plasmon$+$paramagnon-pole approximation to the charge-charge and spin-spin response functions.

From the experimental point of view, as already remarked, electrons in a semiconductor inversion layer or in a quantum well can be modelled by a quasi-$2D$ EG~\cite{ando}. Quantum Shubnikov-de Haas (SdH) oscillations of the magnetoresistance~\cite{abrikosov} provide a powerful tool for measuring Fermi-liquid parameters of a quasi-$2D$ EG. Measurements performed over the past years~\cite{smith,abstreiter,fang,neugebauer} have shown sizeable renormalizations of the QP effective mass and effective Land\'e $g$-factor. These experiments have been performed in a relatively high-density regime, {\it i.e.} for $r_s \lesssim 2$ say. The density dependence of the effective mass $m^*$ was obtained by Smith and Stiles~\cite{smith} from a study of SdH oscillations in Si inversion layers. To obtain the same information Abstreiter {\it et al}.~\cite{abstreiter} used instead cyclotron resonance measurements. Fang and Stiles~\cite{fang} and Neugebauer {\it et al}.~\cite{neugebauer} performed a series of SdH experiments on Si inversion layers and obtained the dependence of the modified Land\'e factor $g^*$ on carrier density. The product of $g^*$ and $m^*$, which is proportional to the spin susceptibility $\chi_S$, can be determined from the SdH oscillations in a tilted magnetic field as suggested in Ref.~\onlinecite{fang}. 

The issue of the apparent metal-insulator transition~\cite{mit} (MIT) in low-density $2D$ electron systems 
has prompted intense experimental studies on quasiparticle properties~\cite{dolgopolov,okamoto,vitkalov,shashkin,pudalov_prl,tutuc,noh,zhu_2003,vakili_2004} 
in the intermediate-to-strong coupling regime, $r_s \gtrsim 2$ say. 
Many authors~\cite{dolgopolov} have shown that the resistance of a Si-MOSFET 
is increased dramatically by increasing the value of an in-plane magnetic field, and saturates at a characteristic value of several Tesla. Performing low-field SdH measurements on Si-MOSFET's, Okamoto {\it et al}.~\cite{okamoto} have shown that the saturation value is the magnetic field that is necessary to fully polarize the electron spins. 
An interpretation~\cite{vitkalov,shashkin} of the in-plane magnetoresistance in Si inversion layers suggested a ferromagnetic instability at or very close to the critical density for the $2D$ MIT driven by a divergence in the effective mass. Direct measurements of $m^*$ in high-mobility Si-MOSFET's  over a wide range of carrier density, using a novel technique based on the beating pattern of SdH oscillations in crossed magnetic fields, have been reported by Pudalov {\it et al}.~\cite{pudalov_prl}. These authors measured $m^*$ and $\chi_S$ in the vicinity of the $2D$ MIT, but found no evidence for a divergent behavior. Only a moderate enhancement of $m^*$ by a factor of $\approx 2-2.5$ over the band mass was observed near the critical density for the $2D$ MIT. Two groups have also reported anomalous density dependences of the Land\'e factor in $n$-doped~\cite{tutuc} ($2\lesssim r_s\lesssim 7$) and $p$-doped~\cite{noh} ($r_s \gtrsim 17$) GaAs/AlGaAs heterojunctions that are in disagreement with results in Si-MOSFET's. The dependence of the spin susceptibility on the degree of spin polarization of the sample can account for this anomalous behavior as pointed out by Zhu {\it et al.}~\cite{zhu_2003}, who studied a $2D$ EG of exceedingly high quality. 

To complete the cornucopia of recent experimental findings on QP properties, it is worth mentioning that Vakili {\it et al.}~\cite{vakili_2004} have reported measurements of $m^*$ and $\chi_S$ in a dilute $2D$ EG confined to a narrow AlAs quantum well (only $45\,{\rm \AA}$ wide). The electron system investigated in Ref.~\onlinecite{vakili_2004} is quite interesting because the electrons occupy an out-of-plane conduction-band valley, rendering the system similar to $2D$ electrons in Si-MOSFET's but with only one valley occupied. Quite surprisingly, the results of Vakili {\it et al.}~\cite{vakili_2004} for $\chi_S$ are in good agreement with the QMC results of Attaccalite {\it et al.}~\cite{attaccalite} even though this simulation has been carried out for a strictly disorder-free EG. This might indicate that $\chi_S$ is not strongly dependent on disorder. On the other hand, there is a significant spread in the experimental results of Ref.~\onlinecite{vakili_2004} for $m^*$, which turns out to be both sample and cool-down dependent. Difficulties associated with the SdH data analysis have been pointed out in Ref.~\onlinecite{vakili_2004} as one of the possible causes for this spread.
 
At this point it is probably worth commenting that indeed there could be in principle 
subtle issues associated with the analysis of the SdH traces in $2D$ systems. In fact the amplitude of the SdH oscillations is usually fitted to the $3D$ Lifshitz-Kosevich (LK) formula~\cite{lk} upon a trivial change in the single-particle spectrum. The fit is based on an impurity scattering Dingle temperature $T_D$ and an ``effective" mass. In recent years a number of caveats concerning the applicability of such a procedure to $2D$ strongly-interacting systems have appeared~\cite{caveats}. In particular, Martin {\it et al}.~\cite{maslov} have shown that the interplay between electron-electron interactions and electron-impurity scattering leads in $2D$ to an effective temperature-dependent Dingle temperature with a leading low-temperature behavior of the type $T_D(T)\propto T\ln{T}$. The need for the introduction of a temperature-dependent Dingle parameter in strongly-coupled Si-MOSFET's has been emphasized in the above-mentioned Ref.~\cite{pudalov_prl} where a linear $T_D(T)$ was used to fit the longitudinal magnetoresistance data. 
Quantitative differences on the resultant effective mass are found using such type of procedure: roughly speaking, 
the tendency is to get substantially lower values for $m^*$ than those obtained using the same Dingle parameter for all temperatures.

For a quantitative comparison between suitable theories that take into account quasi-$2D$ effects 
(such as finite width of the electron wavefunctions in the confinement direction and valley degeneracies)  
and the experimental results~\cite{smith,fang} for Si-MOSFET's in the weak-coupling regime $r_s\lesssim 2$ we refer the reader to the work of Yarlagadda and Giuliani~\cite{yarlagadda_gstar} and references therein. In this work we will try and carry out a comparison between the theory and the experimental data of Zhu {\it et al.}~\cite{private_zhu} for strongly-interacting electrons ($2 \lesssim r_s\lesssim 6$), occupying a single valley in an exceptionally clean GaAs/AlGaAs quantum well.

The contents of the paper are described briefly as follows.  
In Sect.~\ref{quasiparticle_self_energy} we present in great detail the theoretical background. We proceed in Sect.~\ref{local_fields} to discuss the inputs we have used for our numerical calculations, while 
in Sect.~\ref{numerical_results} we present our main results for the real and imaginary part of the quasiparticle self-energy, the many-body enhancement of the effective mass, and the renormalization constant. Finally, in Sect.~\ref{conclusions} we compare our theory with 
the experimental results of Zhu {\it et al.}~\cite{private_zhu} and report some conclusions. 
In order to make the paper fully self-contained we have included two Appendices which contain a number of helpful details on how we have in practice calculated the QP self-energy.

\section{Theory of the quasiparticle self-energy}
\label{quasiparticle_self_energy}
The aim of this Section is to give a sound theoretical justification to the expression for
the retarded QP self-energy $\Sigma_{\rm \scriptstyle ret}({\bf k},\omega)$ of a $2D$ paramagnetic EG 
that will be used throughout this paper and that we have summarized in Eqs.~(\ref{sx}) and (\ref{ch}) below. 

To fix the notation we start by introducing the Hamiltonian 
for a $2D$ EG confined to an area $S$,
\begin{equation}\label{eg_hamiltonian}
{\mathcal H}_{\rm EG}=\sum_{{\bf k},\sigma}\varepsilon_{\bf k}
{\hat c}^{\dagger}_{{\bf k},\sigma}{\hat c}_{{\bf k},\sigma}+\frac{1}{2 S}\sum_{{\bf q}\neq 0}v_{\bf q}\sum_{{\bf k}_1,\sigma_1}\sum_{{\bf k}_2,\sigma_2}{\hat c}^{\dagger}_{{\bf k}_1+{\bf q},\sigma_1}{\hat c}^{\dagger}_{{\bf k}_2-{\bf q},\sigma_2}{\hat c}_{{\bf k}_2,\sigma_2}{\hat c}_{{\bf k}_1,\sigma_1}\,.
\end{equation}
Here ${\hat c}^{\,\dagger}_{{\bf k},\,\sigma}$ and ${\hat c}_{{\bf k},\, \sigma}$ are fermionic creation and annihilation operators which satisfy canonical anticommutation relations, $\varepsilon_{\bf k}=\hbar^2{\bf k}^2/(2m)$ is the single-particle energy, and $v_{\bf q}=2\pi e^2/q$ is the $2D$ Fourier transform of the bare Coulomb interaction $e^2/r$. For later purposes we introduce 
the Fermi wave number $k_F=(2\pi n_{\rm \scriptscriptstyle 2D})^{1/2}=\sqrt{2}/(r_s a_B)$, the Fermi energy $\varepsilon_F=\hbar^2k^2_F/(2m)$ and the quantity $\xi_{\bf k}=\varepsilon_{\bf k}-\varepsilon_F$.

The retarded QP self-energy $\Sigma_{\rm \scriptstyle ret}({\bf k},\omega)$ is written as the sum of two terms,
\begin{equation}\label{inizium_fiat}
\Sigma_{\rm \scriptstyle ret}({\bf k},\omega)=
\Sigma_{\rm \scriptstyle SX}({\bf k},\omega)+\Sigma_{\rm \scriptstyle CH}({\bf k},\omega)
\end{equation}
where the first term is called ``screened-exchange" (SX) and the second term is called ``Coulomb-hole" (CH). The frequency $\omega$ is measured from $\varepsilon_F/\hbar$. 

The SX contribution is given by
\begin{equation}\label{sx}
\Sigma_{\rm \scriptstyle SX}({\bf k},\omega)=
-\int \frac{d^2 {\bf q}}{(2\pi)^2}\,\frac{v_{\bf q}}{
\varepsilon({\bf q},\omega-\xi_{{\bf k}+{\bf q}}/\hbar)}\,\Theta(-\xi_{{\bf k}+{\bf q}}/\hbar)\,.
\end{equation}
Here $\Theta(x)$ is the step function and $\varepsilon({\bf q},\omega)$ is a screening function originating 
from effective Kukkonen-Overhauser interactions~\cite{kukkonen},
\begin{equation}\label{epsilon}
\frac{1}{\varepsilon({\bf q},\omega)}= 1+v_{\bf q}\,\left[1-G_{+}({\bf q},\omega)\right]^2\,\chi_{\rm \scriptstyle C}({\bf q},\omega)+3\,v_{\bf q}\,G^2_{-}({\bf q},\omega)\,\chi_{\rm \scriptstyle S}({\bf q},\omega)\,.
\end{equation}
In Eq.~(\ref{epsilon}) the charge-charge and spin-spin response functions $\chi_{\rm \scriptstyle C}({\bf q}, \omega)$ and $\chi_{\rm \scriptstyle S}({\bf q}, \omega)$ are determined by the spin-symmetric and spin-antisymmetric local-field factors $G_{+}({\bf q},\omega)$ and $G_{-}({\bf q},\omega)$, 
\begin{equation}\label{cc}
\chi_{\rm \scriptstyle C}({\bf q},\omega)
=\frac{\chi_0({\bf q},\omega)}
{1-v_{\bf q}[1-G_+({\bf q},\omega)] \chi_0({\bf q},\omega)}
\end{equation}
and 
\begin{equation}\label{ss}
\chi_{\rm \scriptstyle S}({\bf q},\omega)
=\frac{\chi_0({\bf q},\omega)}{1+v_{\bf q}G_{-}({\bf q},\omega)\,\chi_0({\bf q},\omega)}\,,
\end{equation}
$\chi_0({\bf q},\omega)$ being the Stern response function of a noninteracting $2D$ EG~\cite{stern_1967}. In the paramagnetic electron liquid $G_{\pm}({\bf q},\omega)=[G_{\uparrow\uparrow}({\bf q},\omega)\pm G_{\uparrow\downarrow}({\bf q},\omega)]/2$, where $G_{\sigma\sigma'}({\bf q},\omega)$ are the spin-resolved local fields. Note that $\Sigma_{\rm \scriptstyle SX}({\bf k},\omega)$ is just an ordinary exchange-like self-energy built from the Kukkonen-Overhauser effective interactions instead of bare Coulomb interactions, which would lead to the frequency-independent Hartree-Fock self-energy first calculated for the $2D$ EG by Stern~\cite{stern_1973}.

The CH contribution to the retarded self-energy is given by
\begin{equation}\label{ch}
\Sigma_{\rm \scriptstyle CH}({\bf k},\omega)=
-\int \frac{d^2 {\bf q}}{(2\pi)^2}
\,v_{\bf q}\int_{0}^{+\infty}\,\frac{d \Omega}{\pi}\,\,\frac{\Im m[\varepsilon^{-1}({\bf q},\Omega)]}{\omega-\xi_{{\bf k}+{\bf q}}/\hbar-\Omega+i\delta}\,,
\end{equation}
where $\delta$ is a positive infinitesimal. The real and imaginary part of the retarded self-energy are readily obtained from Eqs. (\ref{sx}) and (\ref{ch}) with the result
\begin{eqnarray}\label{real_sigma}
\Re e\Sigma_{\rm \scriptstyle ret}({\bf k},\omega)=&-&\int \frac{d^2 {\bf q}}{(2\pi)^2}\,
v_{\bf q}\Re e[\varepsilon^{-1}({\bf q},\omega-\xi_{{\bf k}+{\bf q}}/\hbar)]
\,\Theta(-\xi_{{\bf k}+{\bf q}}/\hbar)\\ \nonumber
&-&\int \frac{d^2 {\bf q}}{(2\pi)^2}
\,v_{\bf q}{\mathcal P}\int_{0}^{+\infty}\,\frac{d \Omega}{\pi}\,
\frac{\Im m[\varepsilon^{-1}({\bf q},\Omega)]}{\omega-\xi_{{\bf k}+{\bf q}}/\hbar-\Omega}\,,
\end{eqnarray}
and
\begin{equation}\label{im_sigma}
\Im m\Sigma_{\rm \scriptstyle ret}({\bf k},\omega)
=\int \frac{d^2 {\bf q}}{(2\pi)^2}\,
v_{\bf q}\Im m[\varepsilon^{-1}({\bf q},\omega-\xi_{{\bf k}+{\bf q}}/\hbar)]
\left[\Theta(\omega-\xi_{{\bf k}+{\bf q}}/\hbar)-\Theta(-\xi_{{\bf k}+{\bf q}}/\hbar)\right]\,.
\end{equation}

Once the QP self-energy is known, 
the QP excitation energy $\delta {\mathcal E}_{\rm QP}({\bf k})$, which is the QP energy measured from the 
chemical potential $\mu$ of the interacting EG, can be calculated 
by solving self-consistently the Dyson equation
\begin{eqnarray}\label{dyson_regularized}
\delta {\mathcal E}_{\rm QP}({\bf k})&=&\xi_{\bf k}+\left. 
\Re e \Sigma^{\rm R}_{\rm \scriptstyle ret}({\bf k},\omega)\right|_
{\omega=\,\delta {\mathcal E}_{\rm QP}({\bf k})/\hbar}\,,
\end{eqnarray}
where $\Re e\Sigma^{\rm R}_{\rm \scriptstyle ret}({\bf k},\omega)=
\Re e \Sigma_{\rm \scriptstyle ret}({\bf k},\omega)-\Sigma_{\rm \scriptstyle ret}(k_F,0)$. For later purposes we introduce at this point the so-called on-shell approximation (OSA). This amounts to approximating the QP excitation energy by calculating $\Re e \Sigma^{\rm R}_{\rm \scriptstyle ret}({\bf k},\omega)$ in Eq.~(\ref{dyson_regularized}) at the frequency $\omega=\xi_{\bf k}/\hbar$ corresponding to the single-particle energy, that is
\begin{equation}\label{osa}
\delta {\mathcal E}_{\rm QP}({\bf k})\simeq \xi_{\bf k}+\left. \Re e \Sigma^{\rm R}_{\rm \scriptstyle ret}({\bf k},\omega)\right|_
{\omega=\xi_{\bf k}/\hbar}\,.
\end{equation}

In Sects.~\ref{NgDerivation} and \ref{rha} 
we will give a formal justification of Eqs.~(\ref{sx}) and (\ref{ch}) using two completely different methods: diagrammatic perturbation theory (Sect.~\ref{NgDerivation}) and renormalized Hamiltonian approach (Sect.~\ref{rha}). In the following we shall soon drop, however, the frequency dependence of the local-field factors. 
Recent studies~\cite{nifosi} have evaluated it in the long-wavelength limit ${\bf q}\rightarrow 0$, but the knowledge of the full dependence on wave number is necessary for the type of calculations that we are interested in.

\subsection{Diagrammatic perturbation theory}
\label{NgDerivation}
In this Section we  use a diagrammatic approach that was first developed by Ng and Singwi~\cite{ng}, building on earlier ideas by Vignale and Singwi~\cite{vignale_1985}.
The starting point is the identity~\cite{nozieres}
\begin{equation}
\label{ng1}
\delta \Sigma_\sigma({\bf k},\omega) = i\sum_{\sigma'} \int\frac{d^2{\bf k}'d\omega'}{(2 \pi)^3}  I_{{\bf k}\omega \sigma, {\bf k}' \omega' \sigma'}(0)
\delta G_{\sigma'} ({\bf k}',\omega')~,
\end {equation}
where $\delta \Sigma_\sigma ({\bf k},\omega)$ and  $\delta G_{\sigma}({\bf k},\omega)$ are infinitesimal changes in the self-energy and the Green's function, and $I_{{\bf k}\omega \sigma, {\bf k}'
\omega'\sigma'}(0)$ is the  irreducible electron-hole interaction at zero momentum and energy transfer. 
This identity is graphically represented in Fig.~1. 
The defining feature  of the irreducible electron-hole scattering block $I$ is that it includes only diagrams that cannot be divided  into two parts by cutting a single electron-hole pair propagator  carrying zero energy and momentum.
\begin{figure}
\begin{center}
\includegraphics[scale=0.6]{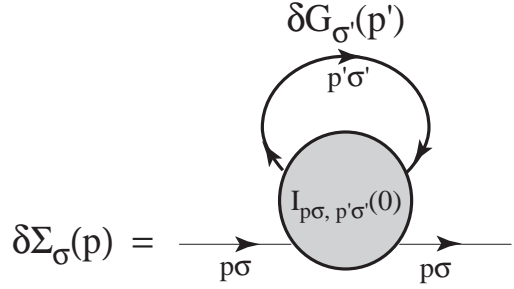}
\caption{Diagrammatic representation of the identity~(\ref{ng1}). Here and in the following we use the four-momentum variable $p$ as a shorthand for $({\bf k},\omega)$.}
\end{center}
\label{WardIdentity}
\end{figure}

The differential relation~(\ref{ng1}) cannot be  integrated as it stands because $I$ is a complicated functional of $G$. The idea of  Ng and Singwi was to use an approximate form of $I_{{\bf k}\omega \sigma, {\bf k}'
\omega'\sigma'}(0)$  that does not depend on $G$.
The ``local approximation" introduced by Vignale and Singwi in their study of the effective electron-electron interaction~\cite{vignale_1985} is useful for this purpose since it yields by physical arguments an expression of the form
\begin{equation}
\label{ng2}
I_{{\bf k}\omega \sigma, {\bf k}' \omega'\sigma'}(0)  \simeq
V^{\rm eff}_{\sigma\sigma'}({\bf k}-{\bf k}',\omega - \omega')~,
\end {equation}
where $V^{\rm eff}_{\sigma\sigma'}$ 
is just a function of the momentum and energy transfers in the electron-hole channel. 
Thus the main characteristic of the Ng-Singwi approach is
that the {\it key approximation in Eq.~(\ref{ng2}) is made
on the irreducible electron-hole interaction rather than on the
self-energy itself}.
With this approximation we can integrate Eq.~(\ref{ng1})
and obtain, up to an integration constant, the result
\begin{equation}
\label{ng3}
 \Sigma_{\sigma} ({\bf k},\omega)=i\sum_{\sigma'} \int \frac{d^{2}{\bf k}' d\omega'}{(2 \pi)^{3}} V^{\rm eff}_{\sigma\sigma'}({\bf k}-{\bf k}',\omega - \omega')
G_{\sigma'}({\bf k}',\omega')~.
\end{equation}
With the replacements ${\bf k}-{\bf k}'={\bf q}$ and $\omega-\omega'=\Omega$, this
expression has the form of the GW approximation~\cite{hedin} except for  two crucial differences: 
(i) the effective interaction $V^{\rm eff}_{\sigma\sigma'}({\bf q},\Omega)$  
includes vertex corrections and is therefore more general than
the screened interaction $W({\bf q},\Omega)$ between test charges that
appears in the GW approximation; and (ii) 
the expression~(\ref{ng3}) involves an undetermined integration
constant which must be fixed by independent means, for example by
requiring that $\Sigma(k_F, 0)$ reproduces the correct value of the
chemical potential as determined from QMC data. An analytic continuation procedure allows one to recast the
time-ordered self-energy in Eq.~(\ref{ng3}) into a retarded self-energy given by the sum of SX and CH contributions as in Eq.~(\ref{inizium_fiat}).

In their derivation of the ``local approximation" in Eq.~(\ref{ng2}) Ng and Singwi sorted  the diagrams that contribute to the irreducible electron-hole interaction into three classes: (1) 
diagrams that are reducible in the  ``crossed"
particle-hole channel, carrying momentum  ${\bf q}$ and energy $\Omega$; (2) 
diagrams that are reducible in the particle-particle channel; and (3) 
diagrams that are irreducible in the particle-particle channel and in both particle-hole channels. The diagrams of  class (1) are shown in Fig.~2. 
\begin{figure}
\begin{center}
\includegraphics[scale=0.6]{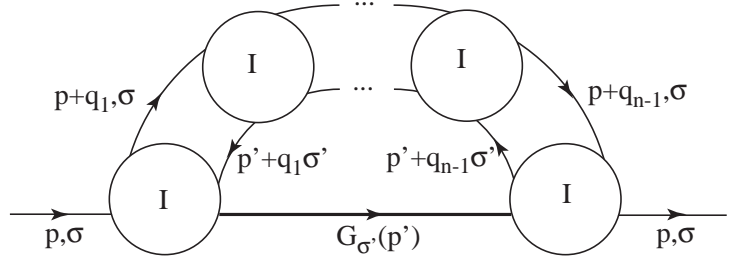}
\caption{Structure of the diagrams included in the approximate
evaluation of  $ I_{p \sigma, p' \sigma'}(0)$. The  number of irreducible
interaction blocks is $n \geq 2$, and there is an integral over the internal four-momenta 
$q_1,...,q_{n-1}$.  Notice that these diagrams are
reducible in the particle-hole channel that carries four-momentum
$p-p'$, but irreducible in the particle-hole channel
defined by the two external legs, which carries zero four-momentum. 
The irreducible interaction blocks are further analyzed in Fig.~3.}
\end{center}
\label{railroaddiagrams.eps}
\end{figure}
Notice that these diagrams are expressed in terms of  three building blocks as shown in Fig.~3,  which are irreducible in the ``crossed" particle-hole channel.   
\begin{figure}
\begin{center}
\includegraphics[scale=0.6]{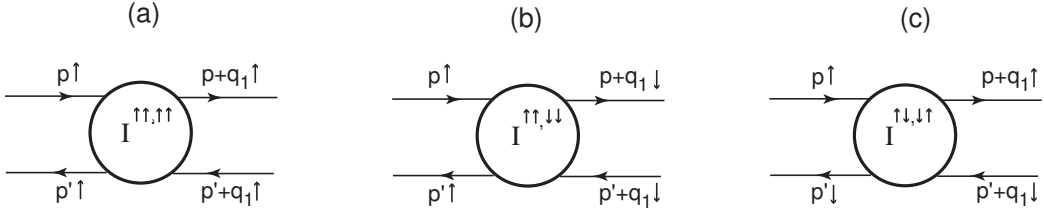}
\caption{The three irreducible blocks out of which  the ``railroad diagrams" of Fig.~2 are constructed.}
\end{center}
\label{blocks.eps}
\end{figure}
In block (a) the particle and the  hole have parallel spin orientations which are conserved as they scatter  against each other.  In block (b) the particle and the  hole have parallel spin orientations, which are reversed as a result of the scattering process.  Finally, in block (c) the particle and the hole have opposite spin orientations. 

In order to make further progress we now  approximate these irreducible blocks by functions of the electron-hole momentum ${\bf q}={\bf k}-{\bf k}'$.  The form of these functions is determined by requiring that the same approximation, when applied to the evaluation of the diagrams for the density-density and spin-spin response functions, yields Eqs.~(\ref{cc}) and~(\ref{ss}). 
For the two diagrams 3(a) and 3(b)  this is accomplished by setting 
\begin{equation}
\label{localapprox}
I^{\sigma\sigma,\tau \tau} \simeq  v_{{\bf q}}[1- G_{\sigma\tau}({\bf q})]~,
\end {equation}
where $\sigma$ and $\tau$ are the spin orientations of the particle and of the hole before and after scattering (the labelling is explained in Fig.~3).  For  diagram 3(c) a simple reasoning based on isotropy in spin space for 
the paramagnetic state leads to 
\begin{equation}
\label{localapprox2}
I^{\sigma\bar\sigma, \bar\sigma\sigma}=I^{\sigma\sigma, \sigma\sigma}-I^{\sigma\sigma, \bar\sigma\bar\sigma}
\simeq-2 v_{{\bf q}}G_-({\bf q})~,
\end {equation}
as one can see from Fig. 4.
\begin{figure}
\begin{center}
\includegraphics[scale=0.6]{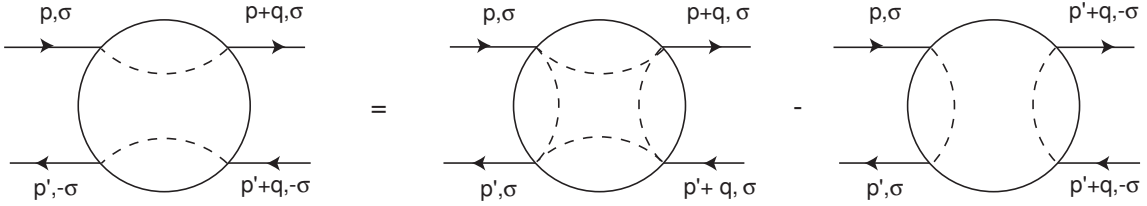}
\caption{Graphical illustration of the relationship between different components of the irreducible electron-hole scattering amplitude in a spin-unpolarized state.}
\end{center}
\label{direct-exch-diags}
\end{figure}

The integrals over the internal four-momenta in Fig.~2 can be carried out analytically if the intermediate Green's functions are replaced by noninteracting Green's functions, giving a result proportional to $[\chi_0(p-p')]^n$ where $\chi_0$ is the Stern function. The whole series of diagrams can then be summed algebraically yielding
\begin{eqnarray}
\label{sum1}
V^{{\rm eff},(1)}_{\sigma\sigma'}({\bf q},\Omega) &=&  [v_{{\bf q}}
G_{+}({\bf q})]^2 \chi_{C}({\bf q},\Omega)\delta_{\sigma,\sigma'}+[v_{{\bf q}}
G_{-}({\bf q})]^2 \chi_{S}({\bf q},\Omega)\delta_{\sigma,\sigma'}\nonumber\\
&+&2 [v_{{\bf q}}
G_{-}({\bf q})]^2 \chi_{S}({\bf q},\Omega)   \delta_{\sigma,-\sigma'}~.  
\end{eqnarray}
The subscripts $+$ and $-$ denote as usual the charge and the
longitudinal spin channels.  The density-density and
spin-spin response functions are expressed in terms of the Stern function
and of the many-body local field factors according to Eqs.~(\ref{cc})~and~(\ref{ss}).  

The evaluation of the remaining terms, {\it i.e.} the particle-particle ladder
diagrams and the fully irreducible diagrams, is considerably more complex. The
only simple diagram is the bare exchange interaction with momentum transfer ${\bf q}$, which 
yields the Hartree-Fock self-energy. All other terms are dropped in the possibly naive hope that they
are smaller than the terms retained in Eq.~(\ref{sum1}).
Keeping only the bare exchange interaction and combining it with
Eq.~(\ref{sum1}) we arrive at an effective interaction of the
form
\begin{equation}\label{ultima}
\left\{
\begin{array}{l}
V^{\rm eff}_{\uparrow\uparrow}({\bf q},\Omega)=
v_{\bf q}+\{v_{\bf q}[1-G_+({\bf q})]\}^2\chi_{C}({\bf q},\Omega)+[v_{\bf q}G_-({\bf q})]^2\chi_{S}({\bf q},\Omega)\\
V^{\rm eff}_{\uparrow\downarrow}({\bf q},\Omega)=2[v_{\bf q}G_-({\bf q})]^2\chi_{S}({\bf q},\Omega)~.
\end{array}
\right.
\end{equation}
Inserting this in Eq.~(\ref{ng3}) and repeating standard analytical transformations~\cite{hedin} 
one easily recovers the expressions~(\ref{sx}) and~(\ref{ch}) 
for the screened-exchange and Coulomb-hole contributions to the self-energy.

Let us emphasize again that the result that we have obtained by a diagrammatic method rests on Eq.~(\ref{ng3}) for the self-energy (apart from an additive constant) and on the use of the effective interactions shown in Eq.~(\ref{ultima}).
It should be evident from the above derivation that no diagram for
$I_{{\bf k} \omega \sigma, {\bf k}' \omega'\sigma'}(0)$  has been double-counted. Rather, many diagrams have
been dropped, but the result for the self-energy remains to be adjusted {\it a posteriori} by fixing the addivite constant through the correct value of the chemical potential.

Let us see, on the other hand, what would have happened if we had
applied the local approximation directly to the 
self-energy starting from the alternative exact expression,
\begin{equation}
\label{crit2}
\Sigma_{\sigma}({\bf k},\omega) = i \sum_{\sigma'}
\int \frac{d^{2}{\bf k}' d \omega'}{(2 \pi)^{3}}
\frac{v_{{\bf k} - {\bf k}'}}{\epsilon({\bf k}-{\bf k}',\omega-\omega')} 
\tilde \Lambda_{{\bf k} \omega, {\bf k}' \omega'}
G_{\sigma'}({\bf k}',\omega')~,
\end {equation}
where $\tilde \Lambda$ is the proper vertex function and $\epsilon$ is the regular dielectric function (see Fig. 5). Within the local approximation one finds~\cite{rice,rahman_1984,mahan_1989} 
\begin{equation}
\label{localvertex}
\tilde \Lambda_{{\bf k} \omega, {\bf k}' \omega'}=
\frac{1}{1+v_{{\bf k}-{\bf k}'}G_+({\bf k}-{\bf k}') \chi_0({\bf k}-{\bf k}',\omega-\omega')}~,
\end{equation}
so that this route to the self-energy includes only the contribution of charge fluctuations but misses completely that of spin fluctuations. The root of the difficulty obviously lies in the fact that the local approximation
for the vertex function is not good enough to capture the contribution of
spin-density fluctuations. On the other hand, the dependence of $I$ on
spin fluctuations is manifest in the terms proportional to $G_-^2$
in Eq.~(\ref{sum1}). This is the main physical reason why it is
better to apply the local approximation to the differential
relation~(\ref{ng1}) than to the integral
relation~(\ref{crit2}). In fact, all quasiparticle properties of our present interest depend on relative variations of the self-energy, {\it i.e.} on $\delta \Sigma$ rather than on the absolute value of $\Sigma$.

\begin{figure}
\begin{center}
\includegraphics[scale=0.6]{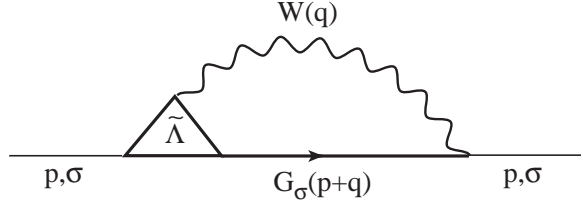}
\caption{Diagrammatic representation of Eq.~(\ref{crit2}). Here $W({\bf q},\Omega)=v_{\bf q}/\epsilon({\bf q},\Omega)$ is the usual test charge-test charge screened interaction.}
\end{center}
\label{WardIdentity.eps}
\end{figure}

\subsection{Renormalized Hamiltonian approach}
\label{rha}
In this Section we discuss 
the derivation of Eqs.~(\ref{sx}) and~(\ref{ch}) from the point of view of an effective 
renormalized Hamiltonian for the low-energy degrees of freedom of the electron liquid~\cite{yarlagadda_1989,yarlagadda_1994,yarlagadda_gstar}. 

We start by dividing the Hilbert space of the EG Hamiltonian ${\mathcal H}_{\rm EG}$ reported in Eq.~(\ref{eg_hamiltonian}) into a ``slow" sector (${\mathcal S}_{\Lambda}$) and a ``fast"  sector (${\mathcal F}_{\Lambda}$), assuming the existence of the Fermi surface at $k=k_F$. ${\mathcal S}_{\Lambda}$ contains only plane-wave states with wavevector ${\bf k}$ close to the Fermi surface, {\it i.e.} such that $|k-k_F|<\Lambda$ 
where $\Lambda$ is an arbitrarily small cutoff. ${\mathcal F}_{\Lambda}$ contains all the other states. We correspondingly introduce ``slow" and ``fast" creation and annihilation operators which operate in these two sectors,
\begin{equation}
{\hat c}_{{\bf k},\sigma}=\left\{
\begin{array}{l}
{\hat s}_{{\bf k},\sigma},\,\, {\bf k}\in {\mathcal S}_{\Lambda}\,,\\
{\hat f}_{{\bf k},\sigma},\,\, {\bf k}\in {\mathcal F}_{\Lambda}\,.
\end{array}
\right.  
\end{equation}
Our aim is to derive an effective Hamiltonian for the slow sector which contains only the ${\hat s}_{{\bf k},\sigma}$ operators by integrating out in a reasoned manner the ${\hat f}_{{\bf k},\sigma}$ degrees of freedom. 

${\mathcal H}_{\rm EG}$ is first rewritten using the 
${\hat s}_{{\bf k},\sigma}$ and ${\hat f}_{{\bf k},\sigma}$ operators,
\begin{equation}\label{eg_different}
{\mathcal H}_{\rm EG}={\mathcal H}_{\rm s}+{\mathcal H}_{\rm f}+{\mathcal H}_{\rm sf}\,.
\end{equation}
The first term is 
\begin{equation}
{\mathcal H}_{\rm s}=\sum_{{\bf k},\sigma}\varepsilon_{\bf k}{\hat s}^{\dagger}_{{\bf k},\sigma}{\hat s}_{{\bf k},\sigma}+\frac{1}{2 S}\sum_{{\bf q}\neq 0}v_{\bf q}\sum_{{\bf k}_1,\sigma_1}\sum_{{\bf k}_2,\sigma_2}{\hat s}^{\dagger}_{{\bf k}_1+{\bf q},\sigma_1}{\hat s}^{\dagger}_{{\bf k}_2-{\bf q},\sigma_2}{\hat s}_{{\bf k}_2,\sigma_2}{\hat s}_{{\bf k}_1,\sigma_1}\,,
\end{equation}
where all wavevectors belong to ${\mathcal S}_{\Lambda}$. Note that the second term in ${\mathcal H}_{\rm s}$, 
which represents the direct interaction between slow particles,
tends to zero faster than the kinetic energy term in the limit $\Lambda\rightarrow 0$. 
Thus this term will be treated by first-order perturbation theory below.

The second term in Eq.~(\ref{eg_different}) 
is the Hamiltonian for the fast sector,
\begin{equation}
{\mathcal H}_{\rm f}=\sum_{{\bf k},\sigma}\varepsilon_{\bf k}
{\hat f}^{\dagger}_{{\bf k},\sigma}{\hat f}_{{\bf k},\sigma}+\frac{1}{2 S}\sum_{{\bf q}\neq 0}v_{\bf q}\sum_{{\bf k}_1,\sigma_1}\sum_{{\bf k}_2,\sigma_2}{\hat f}^{\dagger}_{{\bf k}_1+{\bf q},\sigma_1}{\hat f}^{\dagger}_{{\bf k}_2-{\bf q},\sigma_2}{\hat f}_{{\bf k}_2,\sigma_2}{\hat f}_{{\bf k}_1,\sigma_1}
\end{equation}
where all wavevectors belong to ${\mathcal F}_{\Lambda}$. ${\mathcal H}_{\rm f}$ tends to the full EG Hamiltonian in the limit $\Lambda\rightarrow 0$ and this property will be useful in what follows.

The third term ${\mathcal H}_{\rm sf}$ 
describes the interaction between the slow and the fast particles: this term is the sum of 
fourteen different terms, but here we assume that the relevant operators 
are those which separately conserve the number of particles 
in the two sectors, {\it i.e.} terms of the type ${\hat s}^{\dagger}_{{\bf k}_1+{\bf q},\sigma_1}{\hat s}^{\dagger}_{{\bf k}_2-{\bf q},\sigma_2}{\hat s}_{{\bf k}_2,\sigma_2}{\hat f}_{{\bf k}_1,\sigma_1}$ (trilinear in the field operators of either slow or fast particles) or ${\hat s}^{\dagger}_{{\bf k}_1+{\bf q},\sigma_1}{\hat s}^{\dagger}_{{\bf k}_2-{\bf q},\sigma_2}{\hat f}_{{\bf k}_2,\sigma_2}{\hat f}_{{\bf k}_1,\sigma_1}$ will be dropped. Four terms are left within this assumption,
\begin{eqnarray}
{\mathcal H}_{\rm sf}&\simeq&\frac{1}{2S}\sum_{{\bf q}\neq 0}v_{\bf q}
\sum_{{\bf k}_1,\sigma_1}\sum_{{\bf k}_2,\sigma_2}
{\hat s}^{\dagger}_{{\bf k}_1+{\bf q},\sigma_1}
{\hat f}^{\dagger}_{{\bf k}_2-{\bf q},\sigma_2}{\hat f}_{{\bf k}_2,\sigma_2}{\hat s}_{{\bf k}_1,\sigma_1}
\nonumber\\
&+&\frac{1}{2S}\sum_{{\bf q}\neq 0}v_{\bf q}
\sum_{{\bf k}_1,\sigma_1}\sum_{{\bf k}_2,\sigma_2}
{\hat f}^{\dagger}_{{\bf k}_1+{\bf q},\sigma_1}
{\hat s}^{\dagger}_{{\bf k}_2-{\bf q},\sigma_2}{\hat s}_{{\bf k}_2,\sigma_2}{\hat f}_{{\bf k}_1,\sigma_1}\nonumber\\
&+&\frac{1}{2S}\sum_{{\bf q}\neq 0}v_{\bf q}
\sum_{{\bf k}_1,\sigma_1}\sum_{{\bf k}_2,\sigma_2}
{\hat s}^{\dagger}_{{\bf k}_1+{\bf q},\sigma_1}{\hat f}^{\dagger}_{{\bf k}_2-{\bf q},\sigma_2}{\hat s}_{{\bf k}_2,\sigma_1}{\hat f}_{{\bf k}_1,\sigma_2}\nonumber\\
&+&\frac{1}{2S}\sum_{{\bf q}\neq 0}v_{\bf q}
\sum_{{\bf k}_1,\sigma_1}\sum_{{\bf k}_2,\sigma_2}
{\hat f}^{\dagger}_{{\bf k}_1+{\bf q},\sigma_1}{\hat s}^{\dagger}_{{\bf k}_2-{\bf q},\sigma_2}{\hat f}_{{\bf k}_2,\sigma_1}{\hat s}_{{\bf k}_1,\sigma_2}
\end{eqnarray}
where ${\bf k}_1$ and ${\bf k}_1+{\bf q}$ belong to ${\mathcal S}_{\Lambda}$, and  ${\bf k}_2$ and ${\bf k}_2-{\bf q}$
belong to ${\mathcal F}_{\Lambda}$. By simple 
algebraic manipulations ${\mathcal H}_{\rm sf}$ can be written as
\begin{eqnarray}
{\mathcal H}_{\rm sf}&=&\frac{1}{S}\sum_{{\bf q}\neq 0}\sum_{{\bf k}_1,\sigma_1}\sum_{{\bf k}_2,\sigma_2}
v_{\bf q}{\hat s}^{\dagger}_{{\bf k}_1+{\bf q},\sigma_1}{\hat f}^{\dagger}_{{\bf k}_2-{\bf q},\sigma_2}{\hat f}_{{\bf k}_2,\sigma_2}{\hat s}_{{\bf k}_1,\sigma_1}\nonumber\\
&-&\frac{1}{S}\sum_{{\bf q}\neq 0}\sum_{{\bf k}_1,\sigma_1}\sum_{{\bf k}_2,\sigma_2}v_{{\bf k}_2-{\bf k}_1-{\bf q}}{\hat s}^{\dagger}_{{\bf k}_1+{\bf q},\sigma_1}{\hat f}^{\dagger}_{{\bf k}_2-{\bf q},\sigma_2}{\hat f}_{{\bf k}_2,\sigma_1}{\hat s}_{{\bf k}_1,\sigma_2}\,.
\end{eqnarray}
The first (direct) term in ${\mathcal H}_{\rm sf}$ describes the Coulomb interaction between slow and fast particles and can then be expressed in terms of density fluctuations in the two sets of particles. The second (exchange) term describes an exchange process in which a slow particle replaces a fast particle and {\it vice versa}. Note that the two particles can have opposite spins, {\it i.e.} $\sigma_1=-\sigma_2$, and when this is the case a net spin angular momentum is exchanged between slow and fast particles. This in practice means that any attempt to write ${\mathcal H}_{\rm sf}$ in terms of collective variables must also involve an interaction between slow and fast particles mediated by spin fluctuations. 

All these arguments bring us to a second crucial approximation: we treat ${\mathcal H}_{\rm sf}$ in an average sense by writing these microscopic processes in terms of interactions between density and spin-density fluctuations in the two sets of slow and fast particles,
\begin{eqnarray}
{\mathcal H}_{\rm sf}&\simeq &
\frac{1}{S}\sum_{{\bf q}\neq 0}v_{\rm C}({\bf q})\sum_{{\bf k},\sigma_1}{\hat n}_{-{\bf q}}
{\hat s}^{\dagger}_{{\bf k}-{\bf q},\sigma_1}{\hat s}_{{\bf k},\sigma_1}\nonumber\\
&+&\frac{1}{S}\sum_{{\bf q}\neq 0}v_{\rm S}({\bf q})\sum_{{\bf k},\sigma_1,\sigma_2}{\hat {\bf S}}_{-{\bf q}}\cdot
\left({\hat s}^{\dagger}_{{\bf k}-{\bf q},\sigma_1}
\left[\mbox{\small \boldmath$\sigma$}\right]_{\sigma_1\sigma_2}
{\hat s}_{{\bf k},\sigma_2}\right)\,,
\end{eqnarray}
where ${\hat n}_{\bf q}=\sum_{{\bf k},\sigma}
{\hat f}^{\dagger}_{{\bf k}-{\bf q},\sigma}{\hat f}_{{\bf k},\sigma}$ and ${\hat {\bf S}}_{\bf q}=
\sum_{{\bf k},\sigma,\sigma'}{\hat f}^{\dagger}_{{\bf k}-{\bf q},\sigma}
\left[\mbox{\small \boldmath$\sigma$}\right]_{\sigma\sigma'}
{\hat f}_{{\bf k},\sigma'}$ are, respectively, the density and spin-density operators for the fast sector. The effective interaction potentials $v_{\rm C}({\bf q})$ and $v_{\rm S}({\bf q})$ must include both the exchange and the correlations effects. This requirement can be fullfilled in an approximate way by means of local field factors $G_+({\bf q})$ and $G_-({\bf q})$ by taking $v_{\rm C}({\bf q})=v_{\bf q}[1-G_+({\bf q})]$ and $v_{\rm S}({\bf q})=-v_{\bf q}G_-({\bf q})$. Note that we are again using frequency-independent 
local field factors for the reasons already indicated above. 

We now carry out a unitary transformation that eliminates the interaction ${\mathcal H}_{\rm sf}$ between slow and fast particles to second order in its strength~\cite{hamann}. We search for a Hermitian operator ${\hat \Theta}_{\Lambda}$ which maps the original Hamiltonian into a new Hamiltonian ${\mathcal H}'=\exp{(i{\hat \Theta}_{\Lambda})}{\mathcal H}_{\rm EG}\exp{(-i{\hat \Theta}_{\Lambda})}$ having the same eigenvalues 
but transformed eigenfunctions. 
The generator ${\hat \Theta}_{\Lambda}$ is at least of first order in the strength of the interaction between the two sectors. The transformed Hamiltonian can then be expanded according to 
\begin{eqnarray}
{\mathcal H}'&=&{\mathcal H}_{\rm EG}+i[{\hat \Theta}_{\Lambda},{\mathcal H}_{\rm EG}]-\frac{1}{2}
[{\hat \Theta}_{\Lambda},[{\hat \Theta}_{\Lambda},
{\mathcal H}_{\rm EG}]]+...\nonumber\\
&=&{\mathcal H}_{\rm s}+{\mathcal H}_{\rm f}+{\mathcal H}_{\rm sf}+i[{\hat \Theta}_{\Lambda},{\mathcal H}_{\rm s}+{\mathcal H}_{\rm f}]+i[{\hat \Theta}_{\Lambda},{\mathcal H}_{\rm sf}]
-\frac{1}{2}[{\hat \Theta}_{\Lambda},[{\hat \Theta}_{\Lambda},{\mathcal H}_{\rm s}+{\mathcal H}_{\rm f}]]+...\,,
\end{eqnarray}
where we have dropped the commutator 
$[{\hat \Theta}_{\Lambda},[{\hat \Theta}_{\Lambda},{\mathcal H}_{\rm sf}]]$ because it is of at least 
third order. The interaction term ${\mathcal H}_{\rm sf}$ is eliminated by choosing 
${\hat \Theta}_{\Lambda}$ as the solution of the operatorial equation
\begin{equation}\label{theta_lambda}
i[{\hat \Theta}_{\Lambda},{\mathcal H}_{\rm s}+{\mathcal H}_{\rm f}]=-{\mathcal H}_{\rm sf}\,.
\end{equation}
Finally, by averaging over the ground-state $|0\rangle$ of 
the Hamiltonian ${\mathcal H}_{\rm f}$ at energy $E_0$, 
we obtain the effective Hamiltonian for the low-energy degrees of freedom of the electron liquid,
\begin{equation}
{\mathcal H}_{\rm QP}\equiv \langle 0|{\mathcal H}'|0\rangle=E_0+
{\mathcal H}_{\rm s}+
\frac{i}{2}\langle 0|[{\hat \Theta}_{\Lambda},{\mathcal H}_{\rm sf}]|0 \rangle\,.
\end{equation}
Obviously $E_0$ does not play any physical role and will be dropped from now on. 

The operatorial equation (\ref{theta_lambda}) 
can be solved for ${\hat \Theta}_{\Lambda}$ once 
the commutator of ${\hat \Theta}_{\Lambda}$ with the interaction term in ${\mathcal H}_{\rm s}$ is dropped, the justification being that this commutator vanishes upon averaging on the ground-state of the fast sector.
A lengthy but straightforward calculation yields
\begin{eqnarray}\label{h_qp_nno}
{\mathcal H}_{\rm QP}&=&{\mathcal H}_{\rm s}+\frac{1}{2S}
\sum_{{\bf q}\neq 0}v^2_{\rm C}({\bf q})\sum_{{\bf k}_1,\sigma_1}\sum_{{\bf k}_2,\sigma_2}
M_{\rm C}({\bf k}_1,{\bf k}_2,{\bf q})
{\hat s}^{\dagger}_{{\bf k}_1+{\bf q},\sigma_1}{\hat s}_{{\bf k}_1,\sigma_1}
{\hat s}^{\dagger}_{{\bf k}_2-{\bf q},\sigma_2}{\hat s}_{{\bf k}_2,\sigma_2}\nonumber\\
&+&\frac{1}{2S}
\sum_{{\bf q}\neq 0}v^2_{\rm S}({\bf q})\sum_{{\bf k}_1,\sigma_1}\sum_{{\bf k}_2,\sigma_2}
M_{\rm S}({\bf k}_1,{\bf k}_2,{\bf q})
{\hat s}^{\dagger}_{{\bf k}_1+{\bf q},\sigma_1}[\mbox{\small \boldmath$\sigma$}]_{\sigma_1\sigma_2}
{\hat s}_{{\bf k}_1,\sigma_1}\nonumber\\
&\cdot& \left(\sum_{\tau_1,\tau_2}
{\hat s}^{\dagger}_{{\bf k}_2-{\bf q},\tau_1}[\mbox{\small \boldmath$\sigma$}]_{\tau_1\tau_2}{\hat s}_{{\bf k}_2,\tau_2}\right)\,,
\end{eqnarray}
where we have defined
\begin{equation}\label{m_c}
M_{\rm C}({\bf k}_1,{\bf k}_2,{\bf q})=\frac{1}{S}\sum_n
\left[\frac{|\langle n|{\hat n}_{\bf q}|0\rangle|^2}{\varepsilon_{{\bf k}_2}-\varepsilon_{{\bf k}_2-{\bf q}}+E_0-E_n}-\frac{|\langle n|{\hat n}_{\bf q}|0\rangle|^2}{\varepsilon_{{\bf k}_1}-\varepsilon_{{\bf k}_1+{\bf q}}+E_n-E_0}\right]\,,
\end{equation}
and
\begin{equation}\label{m_s}
M_{\rm S}({\bf k}_1,{\bf k}_2,{\bf q})=\frac{1}{S}\sum_n
\left[\frac{|\langle n|{\hat S}_{z,{\bf q}}|0\rangle|^2}{\varepsilon_{{\bf k}_2}-\varepsilon_{{\bf k}_2-{\bf q}}+E_0-E_n}-\frac{|\langle n|{\hat S}_{z,{\bf q}}|0\rangle|^2}{\varepsilon_{{\bf k}_1}-\varepsilon_{{\bf k}_1+{\bf q}}+E_n-E_0}\right]\,.
\end{equation}
Here $|n\rangle$ is an exact excited eigenstate of the fast-sector Hamiltonian with eigenvalue $E_n$. 

A key point is that ${\mathcal H}_{\rm QP}$ as written in Eq.~(\ref{h_qp_nno}) is not normal-ordered with respect to the vacuum, {\it i.e.} contains a self-interaction term that needs to be subtracted. The physical meaning of this term is clear: a slow particle moving through the ``medium" of fast-moving particles creates a local polarization which in turn acts back on it. To subtract this self-interaction term we just proceed 
by normal ordering ${\mathcal H}_{\rm QP}$ with respect to the vacuum, with the result
\begin{eqnarray}\label{h_qp_no}
{\mathcal H}_{\rm QP}&=&
\sum_{{\bf k},\sigma}\left[
\varepsilon_{\bf k}-\frac{1}{S}\sum_{\bf q}\int_0^{+\infty}\frac{d\omega'}{\pi}\,\frac{v^2_{\rm C}({\bf q})\Im m \chi_{\rm C}({\bf q},\omega')+3v^2_{\rm S}({\bf q})\Im m \chi_{\rm S}({\bf q},\omega')}
{\Delta_{{\bf k},{\bf q}}-\omega'}\right]
{\hat s}^{\dagger}_{{\bf k},\sigma}{\hat s}_{{\bf k},\sigma}\nonumber \\
&+&\frac{1}{2S}
\sum_{{\bf q}\neq 0}[v_{\bf q}+v^2_{\rm C}({\bf q})\chi_{\rm C}({\bf q},\Delta_{{\bf k},{\bf q}})]
\sum_{{\bf k}_1,\sigma_1}\sum_{{\bf k}_2,\sigma_2}
{\hat s}^{\dagger}_{{\bf k}_1+{\bf q},\sigma_1}
{\hat s}^{\dagger}_{{\bf k}_2-{\bf q},\sigma_2}
{\hat s}_{{\bf k}_2,\sigma_2}
{\hat s}_{{\bf k}_1,\sigma_1}\nonumber\\
&+&\frac{1}{2S}
\sum_{{\bf q}\neq 0}v^2_{\rm S}({\bf q})\chi_{\rm S}({\bf q},\Delta_{{\bf k},{\bf q}})
\sum_{{\bf k}_1,\sigma_1}\sum_{{\bf k}_2,\sigma_2}\sum_{\tau_1,\tau_2}
\left[\mbox{\small \boldmath$\sigma$}\right]_{\sigma_1\sigma_2}\cdot \left[\mbox{\small \boldmath$\sigma$}\right]_{\tau_1\tau_2}
{\hat s}^{\dagger}_{{\bf k}_1+{\bf q},\sigma_1}
{\hat s}^{\dagger}_{{\bf k}_2-{\bf q},\tau_1}
{\hat s}_{{\bf k}_2,\tau_2}{\hat s}_{{\bf k}_1,\sigma_2}
\,,\nonumber\\
\end{eqnarray}
where $\Delta_{{\bf k},{\bf q}}=(\varepsilon_{\bf k}-\varepsilon_{{\bf k}-{\bf q}})/\hbar$. The sums over the eigenstates of ${\mathcal H}_{\rm f}$ in Eqs.~(\ref{m_c}) and (\ref{m_s}) have been carried out
using the identity
\begin{equation}
\frac{1}{S}\sum_n
\frac{|\langle n|{\hat n}_{\bf q}|0\rangle|^2}{\varepsilon_{\bf k}-\varepsilon_{{\bf k}-{\bf q}}-(E_n-E_0)}=-\frac{1}{\pi}\int_0^{+\infty}d\omega'\frac{\Im m \chi_{\rm C}({\bf q},\omega')}{
(\varepsilon_{\bf k}-\varepsilon_{{\bf k}-{\bf q}})/\hbar-\omega'}
\end{equation}
and a similar identity for ${\hat S}_{z,{\bf q}}$. Here $\chi_{\rm C}({\bf q},\omega)$ and $\chi_{\rm S}({\bf q},\omega)$ are the response functions of the ``medium", which asymptotically tend to those 
of the full EG for $\Lambda\rightarrow 0$. 

The QP Hamiltonian ${\mathcal H}_{\rm QP}$ as written in Eq.~(\ref{h_qp_no}) has a very clear physical meaning. 
It is the Hamiltonian for a gas of weakly interacting slow-moving particles with a single-particle dispersion relation shifted by a self-interaction term (the so-called Coulomb-hole shift) generated by the 
normal ordering with respect to the vacuum. The reason for being weakly interacting is that 
the quartic term in ${\mathcal H}_{\rm QP}$ tends to zero faster than the kinetic-energy term in the limit $\Lambda\rightarrow 0$. We can thus calculate the quasiparticle energy within 
first-order perturbation theory, with the result
\begin{equation}\label{sum_today}
{\mathcal E}_{\rm QP}({\bf k})=\varepsilon_{\bf k}+{\mathcal E}_{\rm SX}({\bf k})+{\mathcal E}_{\rm CH}({\bf k})\,.
\end{equation}
Here the first term is the bare single-particle energy and the second term, which has been generated from the normal ordering with respect to the Fermi sea, is formally an ordinary exchange self-energy calculated with a dynamically screened effective interaction (the Kukkonen-Overhauser QP-QP interaction):
\begin{equation}
{\mathcal E}_{\rm SX}({\bf k})=-\int \frac{d^2{\bf q}}{(2\pi)^2}\left[v_{\bf q}+
v^2_{\rm C}({\bf q})\chi_{\rm C}({\bf q},\Delta_{{\bf k},{\bf q}})+
3v^2_{\rm S}({\bf q})\chi_{\rm S}({\bf q},\Delta_{{\bf k},{\bf q}})\right]\Theta(-\xi_{{\bf k}-{\bf q}}/\hbar)\,.
\end{equation}
The last term in Eq.~(\ref{sum_today}) is the Coulomb-shift generated by the normal ordering with respect to the vacuum,
\begin{equation}
{\mathcal E}_{\rm CH}({\bf k})=-\int \frac{d^2{\bf q}}{(2\pi)^2}
\int_0^{+\infty}\frac{d\omega'}{\pi}\,\frac{v^2_{\rm C}({\bf q})\Im m \chi_{\rm C}({\bf q},\omega')+3v^2_{\rm S}({\bf q})\Im m \chi_{\rm S}({\bf q},\omega')}
{\Delta_{{\bf k},{\bf q}}-\omega'}\,.
\end{equation}
The sum of ${\mathcal E}_{\rm SX}({\bf k})$ and ${\mathcal E}_{\rm CH}({\bf k})$ coincides with Eq.~(\ref{real_sigma}) calculated at the single-particle frequency $\omega=\xi_{\bf k}/\hbar$.

\section{Local field factors}
\label{local_fields}

As is clear from Eqs.~(\ref{inizium_fiat})-(\ref{im_sigma}), the local-field factors are fundamental quantities for an evaluation of quasiparticle properties. In this Section we introduce the static values of these functions, that 
we have chosen to calculate the real and imaginary parts of the 
QP self-energy reported in Eqs.~(\ref{real_sigma}) and (\ref{im_sigma}).

Analytical expressions are available~\cite{davoudi_2001} for $G_+({\bf q})$ and $G_-({\bf q})$, which 
reproduce the most recent Diffusion Monte Carlo data~\cite{moroni_1992,senatore_1999} and, as we are going to summarize below, embody the exact asymptotic behaviors at both small and large wave number $q$. Specifically, in the long wavelength limit our choice satisfies 
the compressibility and spin-susceptibility sum rules,
\begin{equation}\label{smallq}
\lim_{q \rightarrow 0}~G_{\pm}({\bf q})= A_{\pm}\frac{q}{k_{F}}
\end{equation}
with $A_+=(1-\kappa_{0}/\kappa)/(r_{s}\sqrt{2})$ and $A_-=(1-\chi_{P}/\chi_S)/(r_{s}\sqrt{2})$. Here 
$\kappa_{0}$ is the compressibility of the
noninteracting gas, $\kappa$ and $\chi_S$ are the compressibility and the spin susceptibility 
of the interacting system, and $\chi_P$ is the Pauli spin susceptibility.
By making use of the thermodynamic definitions of $\kappa$ and $\chi_S$ we can write
\begin{equation}
\frac{\kappa_{0}}{\kappa}=1-\frac{\sqrt{2}}{\pi}\,r_{s}+\frac{r^{4}_{s}}{8}
\left[\frac{\partial^{2} \varepsilon_{c}(r_{s},0)}{\partial r^{2}_{s}}-\frac{1}{r_{s}}
\frac{\partial \varepsilon_{c}(r_{s},0)}{\partial r_{s}}\right]\,,
\end{equation}
and
\begin{equation}
\frac{\chi_{P}}{\chi_S}=1-\frac{\sqrt{2}}{\pi}\,r_s
+\frac{r^{2}_s}{2}\,\left.\frac{
\partial^{2} \varepsilon_{c}(r_s,\zeta)}{\partial \zeta^{2}}\right|_{\zeta=0}\,, 
\end{equation}
where $\varepsilon_{c}(r_{s},\zeta)$ is the correlation energy per particle as a function of $r_s$ and of the degree $\zeta$ of spin-polarization~\cite{rapisarda,attaccalite}. 

At large $q$, on the other hand, the local fields of Ref.~\onlinecite{davoudi_2001} satisfy the asymptotic behavior~\cite{santoro_holas_vignale,footnote}  
\begin{equation}\label{largeq}
G_{\pm}({\bf q})\rightarrow C\, \frac{q}{k_{F}}+B_\pm\,.
\end{equation}
Here $C$ is determined by the difference in kinetic energy between the
interacting and the ideal Fermi gas,
\begin{equation}
C=-\frac{r_s}{2\sqrt{2}}\,\frac{\partial}{\partial r_s}[r_s \varepsilon_{c}(r_s,0)]\,,
\end{equation}
while $B_+=1-g(0)$ and $B_-=g(0)$, with $g(0)$ being the value of the pair distribution function at the origin~\cite{isawa}.

\section{Numerical results}
\label{numerical_results}
We turn to a presentation of our main numerical results. In Sect.~\ref{qp_energy} we present 
some illustrative results for the QP excitation energy and lifetime, and 
in Section~\ref{mstar} we give our results for the QP effective mass and renormalization constant. 
In all figures the labels ``RPA", ``$G_+$" and ``$G_+ \& G_-$" refer to three possible choices for the local-field factors: ``RPA" refers to the case in which local-field factors are not included, ``$G_+$" to the case in which the antisymmetric spin-spin local field is set to zero ({\it i.e.} spin-density fluctuations are not allowed), and finally ``$G_+ \& G_-$" refers to the full theory including both charge- and spin-density fluctuations.
\subsection{Quasiparticle self-energy}
\label{qp_energy}
We have computed the real and imaginary parts of the QP self-energy 
using Eqs.~(\ref{real_sigma}) and (\ref{im_sigma}). 
In Figs.~6 and~7 we show the real part of the SX and CH contributions as from Eqs.~(\ref{sx}) and~(\ref{ch}), evaluated at the single-particle frequency  $\omega=\xi_{\bf k}/\hbar$ and measured from their 
value at $k=k_F$. Note the presence of a strong dip in the CH term at a value of $k$ ($k_p$, say) which 
depends on $r_s$ and on the functional form of the charge-charge local field factor. This is the plasmon dip, which is also present in $3D$ and originates from the fact that at each $r_s$ there is a sufficiently high value of $k$ for decay of an electron-hole pair into a plasmon with conservation of momentum and energy. Mathematically this dip arises for the reasons explained in Appendix A. 

In Fig.~8 we show $\Re e\Sigma^{\rm R}_{\rm \scriptstyle ret}({\bf k},\omega)$ as from Eq.~(\ref{real_sigma}), evaluated at $\omega=\xi_{\bf k}/\hbar$. There is substantial cancellation between the SX and CH contributions for $k<k_F$,  so that in this range the QP self-energy is  essentially 
very weakly momentum-dependent. Such a function has a Fourier transform which is to a good extent local in real space, and this result can be viewed as a microscopic justification of the local-density approximation to the exchange-correlation potential of density-functional theory. 

In Fig.~9 we show the absolute value of $\Im m\Sigma_{\rm \scriptstyle ret}({\bf k},\omega)$ as from Eq.~(\ref{im_sigma}), evaluated at $\omega=\xi_{\bf k}/\hbar$. This function takes a finite jump at the wave number of the plasmon dip. The discontinuity is peculiar to $2D$~\cite{giuliani_quinn}: it is absent in $3D$ and arises from the fact that the oscillator strength of the plasmon pole is non-zero at $k_p$ (see Appendix A). The qualitative difference in the shape of the imaginary part of the self-energy below and above $k_p$ reflects the opening of a new decay channel for an electron-hole pair.

\subsection{Many-body effective mass enhancement}
\label{mstar}
Once the QP excitation energy is known, the effective mass $m^*$ can be calculated by means of the 
relationship
\begin{equation}\label{ms}
\frac{1}{m^*}=\frac{1}{\hbar^2 k_F} \left.\frac{d \delta {\mathcal E}_{\rm QP}(k)}{dk}\right|_{k=k_F}\,.
\end{equation}
In Sect.~\ref{quasiparticle_self_energy} we remarked that the QP excitation energy may be calculated either by 
solving self-consistently the Dyson equation~(\ref{dyson_regularized}) or by using the OSA in Eq.~(\ref{osa}). In what follows the identity
\begin{equation}\label{derivative}
\frac{d \Re e\Sigma^{\rm R}_{\rm \scriptstyle ret}(k,\omega(k))}{dk}=\left.\partial_k \Re e\Sigma^{\rm R}_{\rm \scriptstyle ret}(k,\omega)\right|_{\omega=\omega(k)}+\left.
\partial_{\omega} \Re e\Sigma^{\rm R}_{\rm \scriptstyle ret}(k,\omega)\right|_{\omega=\omega(k)}
\frac{d\omega(k)}{dk}
\end{equation}
will be used, $\omega(k)$ being an arbitrary function of $k$. 

Using Eqs.~(\ref{ms}) and~(\ref{derivative}) with 
$\omega(k)=\delta {\mathcal E}_{\rm QP}(k)/\hbar$ we find that the effective mass $m^*_{\rm D}$ calculated within the Dyson scheme is given by
\begin{equation}\label{mass_dyson}
\frac{m^*_{\rm D}}{m}=\frac{Z^{-1}}{\displaystyle 1+(m/\hbar^{2} k_F)\left.
\partial_k \Re e \Sigma^{\rm R}_{\rm \scriptstyle ret}(k,\omega)\right|_{k=k_F,\omega=0}}\,.
\end{equation}
The renormalization constant $Z$ that 
measures the discontinuity of the momentum distribution at $k=k_F$ is given by
\begin{equation}
Z=\frac{1}{1-\hbar^{-1} \left.\partial_{\omega} \Re e \Sigma^{\rm R}_{\rm \scriptstyle ret}(k,\omega)\right|_{k=k_F,\omega=0}}\,.
\end{equation}
The normal Fermi-liquid assumption, $0<Z\leq 1$, implies $\left.\partial_{\omega} \Re e \Sigma^{\rm R}_{\rm \scriptstyle ret}(k,\omega)\right|_{k=k_F,\omega=0} \leq 0$.

On the other hand, using Eqs.~(\ref{ms}) and~(\ref{derivative}) with 
$\omega(k)=\xi_{\bf k}/\hbar$ we find that the effective mass $m^*_{\rm OSA}$ within the OSA is given by
\begin{equation}\label{mass_osa}
\frac{m^*_{\rm OSA}}{m}=\frac{1}{\displaystyle 1+(m/\hbar^{2} k_F)\left.
\partial_k \Re e \Sigma^{\rm R}_{\rm \scriptstyle ret}(k,\omega)\right|_{k=k_F,\omega=0}+
\hbar^{-1} \left.\partial_{\omega} \Re e \Sigma^{\rm R}_{\rm \scriptstyle ret}(k,\omega)
\right|_{k=k_F,\omega=0}}\,.
\end{equation}
Of course, Eq.~(\ref{mass_osa}) is a good approximant to the Dyson effective mass only in the weak-coupling limit as can be seen by expanding Eq.~(\ref{mass_dyson}) near the point $r_s=0$. The normal Fermi-liquid assumption with the accompanying inequality otherwise 
implies that a zero of the denominator in Eq.~(\ref{mass_osa}) might occur 
depending on the inputs which are used to calculate the QP self-energy. In fact it is clear from Eq.~(\ref{mass_dyson}) that a singular behavior of the effective mass can only occur if the renormalization constant goes to zero, {\it i.e.} the normal Fermi-liquid assumption breaks down. In this case the singular behavior 
could be interpreted as a quantum phase transition of the $2D$ EG to a non-Fermi-liquid state. 

In Fig.~10 we show our numerical results for $m^*_{\rm D}$ and $m^*_{\rm OSA}$. The effective mass enhancement is substantially smaller in the Dyson-equation calculation than in the OSA, the reason being that a large cancellation occurs between numerator and denominator in Eq.~(\ref{mass_dyson}). In both calculations 
the combined effect of charge and spin fluctuations is to enhance the effective mass over the RPA result, whereas the opposite effect is found if only charge fluctuations are included -- a manifestly incorrect result that neglects the spinorial nature of the electron. For completeness we have also included in Figure~10 the variational QMC results of Kwon {\it et al.}~\cite{kwon_1994}. The reader should bear in mind that the effective mass is not a ground-state property and thus its evaluation by the QMC technique is quite delicate, as it involves the construction 
of excited states. There clearly is quantitative disagreement between our ``best" theoretical results (the ``$G_+ \& G_-/{\rm D}$" predictions) and the QMC data.

In Fig.~11 we show the behavior of the two terms in the denominator of Eq.~(\ref{mass_osa}) as functions of 
$r_s$. This figure clearly shows how a divergence can arise in $m^*_{\rm OSA}$: 
for instance, within the RPA the denominator in Eq.~(\ref{mass_osa}) has a zero at $r_s\simeq 15.5$ 
(see the inset in Fig.~11). Our numerical evidence, within all the three theories 
that we have studied, is that indeed (i) $\partial_\omega \Re e \Sigma^{\rm R}_{\rm \scriptstyle ret}(k_F,0)$ 
is negative as it should for a normal Fermi liquid, and 
monotonically increasing in absolute value as a function of $r_s$; and (ii) 
$\partial_k \Re e \Sigma^{\rm R}_{\rm \scriptstyle ret}(k_F,0)$ 
is positive and monotonically increasing too. Within the theory outlined in this work, which 
uses as a key ingredient the Kukkonen-Overhauser effective screening function in Eq.~(\ref{epsilon}), 
the effect of a charge-only local field is to shift this divergence to higher values of $r_s$, while the opposite occurs upon including both charge and spin fluctuations. For instance, within the ``$G_+ \& G_-/{\rm OSA}$" theory the divergence occurs near $r_s=5$. Within the local approximation of Eq.~(\ref{localvertex}) the situation is different~\cite{asgari_2004} and the effect of a charge-only local field is to shift the divergence to lower values of $r_s$.

In Fig.~12 we show our numerical results for the renormalization constant $Z$ in comparison with 
the QMC data of Ref.~\onlinecite{conti}. The theory underestimates the value of $Z$ over 
the whole range of densities explored. Notice that short-range charge-density fluctuations tend to stabilize 
the normal Fermi liquid, while the simultaneous inclusion 
of charge- and spin-density fluctuations works in the opposite way.

For the sake of completeness we have collected in Table I a summary of our numerical results for $m^*$ and $Z$ at 
a few values of $r_s$. In the next Section we will discuss our results for the QP effective mass in the light of recent experimental results and draw our main conclusions.

\section{Comparison with experimental results and conclusions}
\label{conclusions}
A full analysis of the published data for the effective mass of carriers in 
Si-MOSFET's~\cite{shashkin,pudalov_prl} would require 
a more complete theoretical study, mainly to account for the two-valley nature of the material. 
We will focus here instead on the experimental results of Ref.~\onlinecite{private_zhu} as 
kindly provided to us by Dr. Zhu prior to publication. At present 
the data refer to the range $2\lesssim r_s\lesssim 6$, so that we cannot judge the performance 
of the theory in the weak-coupling regime. A quantitative comparison between theory and experiment 
would also require a refined treatment of a 
series of effects such as those due to disorder and to finite temperature. We restrict our analysis 
to the effect of finite sample thickness, by discussing how a softened Coulomb potential modifies 
$m^*$ against the strictly $2D$ results discussed in Sect.~\ref{mstar} and shown in Fig.~10. The expectation is that the QP effective mass will be noticeably smaller when a softened Coulomb interaction is at work.

We have thus recalculated $m^*$ after renormalizing the bare Coulomb potential by means of a form factor to take into account the finite width of the EG in the GaAs/AlGaAs heterojunction-insulated gate field-effect transistor used in Refs.~\onlinecite{zhu_2003} and~\onlinecite{private_zhu}. 
The appropriate renormalized potential is given by $V_{\bf q}=v_{\bf q}F(qd)/{\bar \kappa}$,
where 
\begin{equation}\label{form_factor}
F(x)=\left(1+\frac{\kappa_{\rm ins}}{\kappa_{\rm sc}}\right)\frac{8+9x+3x^2}{8(1+x)^3}+\left(1-\frac{\kappa_{\rm ins}}{\kappa_{\rm sc}}\right)\frac{1}{2(1+x)^6}\,,
\end{equation}
with $d=[\hbar^2 \kappa_{\rm sc}/(48\pi m_z e^2n^*)]^{1/3}$ representing an effective width of the $2D$ EG~\cite{ando}. Here $\kappa_{\rm ins}=10.9$ and $\kappa_{\rm sc}=12.9$ are the dielectric constants of 
the insulator and of the space charge layer, ${\bar \kappa}$ is their average, $m$ is the band mass in the confinement direction, and $n^*=n_{\rm depl}+11n_{\rm \scriptscriptstyle 2D}/32$, the depletion layer charge density $n_{\rm depl}$ being zero in the experiments of Ref.~\onlinecite{private_zhu}. The results that we obtain with the softened potential are shown in Fig.~13. A 
caveat to keep in mind is that we have used the same local-field factors as for a zero-thickness $2D$ EG 
in the lack of a better choice. Thus the results labelled by ``$G_+$" and ``$G_+\&G_-$" in Fig.~13 contain the effect of finite thickness only through the renormalization of the Coulomb potential. We believe that the explicit dependence of the local fields on the finite width of the $2D$ EG should not change the results of Fig.~13 in a substantial manner. 

Comparing the results of Fig.~13 with those in Ref.~\onlinecite{private_zhu} we can draw the following conclusions: 
(i) the ``$G_+$" results, at both the OSA and the Dyson-equation level, do not have the proper functional shape to 
account for the experimental data; (ii) the RPA and ``$G_+\&G_-$" results are rather similar; and (iii) 
the ``$G_+\&G_-$" results,  which treat charge and spin fluctuations on the same footing, show the best performance against the experimental data. In fact, without the use of any fitting parameters, the ``$G_+\&G_-$" results within the OSA compare in a very reasonable manner with the data. The Dyson-equation results, show instead a relatively small and slowly increasing mass enhancement over the whole range of densities, as discussed for the strictly $2D$ case in Sect.~\ref{mstar}.

In summary, we have revisited the problem of the microscopic
calculation of the quasiparticle self-energy and many-body effective mass 
enhancement in a $2D$ EG. We have performed a systematic study 
based on the many-body local-fields theory, taking advantage of the results of 
the most recent Diffusion Monte Carlo calculations of the static charge and spin response of the EG expressed through static local-field factors. We have carried out extensive calculations of both the real and the imaginary part of the 
quasiparticle self-energy. We have also presented results for the effective mass 
enhancement and for the renormalization constant over a wide range of coupling strength. 
In this respect we have critically examined the merits 
of the on-shell approximation {\it versus} the  Dyson-equation calculation. Depending on the local-field factors, 
the OSA predicts a divergence of the effective mass at strong coupling and 
a solution of the Dyson equation is necessary in order to obtain a 
well behaved effective mass. The comparison with the experimental data of 
Ref.~\onlinecite{private_zhu} shows that the simultaneous inclusion of 
charge- and spin-density fluctuations beyond the Random Phase Approximation is 
crucial in accounting for exchange and short-range correlations, and can lead to substantial corrections at 
low carrier densities. A possible role of dynamic correlations, entering through the frequency dependence of the local-field factors, remains to be examined.

\acknowledgements
This work was partially supported by MIUR through the PRIN2001 and PRIN2003 programs. We are especially indebted to Dr. Zhu for showing us her experimental results prior to publication. 
M.P. thanks Prof. Shayegan, Dr. Tan, Dr. Vakili, and  Dr. Zhu 
for their insightful feedback on the experimental results and Dr. S. Moroni 
for useful discussions about the application of the QMC technique to the calculation of Landau parameters. 
G.V. acknowledges support from NSF Grant No. DMR 0313681.
\newpage

\section*{Appendix A. Details on the explicit calculation of the real part of the CH contribution}
Using Eq.~(\ref{ch}) we find that 
the real part of the CH term evaluated at $\omega=\xi_{\bf k}/\hbar$ is given by
\begin{equation}\label{yg_rech}
\left.\Re e\Sigma_{\rm \scriptstyle CH}({\bf k},\omega)\right|_{\omega=\xi_{\bf k}/\hbar}=
-\int \frac{d^2 {\bf q}}{(2\pi)^2}\,v_{\bf q}{\mathcal P}\int_{0}^{+\infty}\,\frac{d \Omega}{\pi}\,
\frac{\Im m[\varepsilon^{-1}({\bf q},\Omega)]}{\xi_{\bf k}/\hbar-\xi_{{\bf k}+{\bf q}}/\hbar-\Omega}\,.
\end{equation}
The angular integration can be performed analytically, with the result
\begin{equation}\label{angular_complex}
\int_0^{2\pi}\frac{d\theta}{\xi_{\bf k}/\hbar-\xi_{{\bf k}+{\bf q}}/\hbar-\Omega}=
\frac{2\pi\,\Theta(\Omega-\Omega_{\rm min}(k))}{\sqrt{[\Omega+\hbar q^2/(2m)]^2-\hbar^2 k^2q^2/m^2}}
\end{equation}
where $\Omega_{\rm min}(k)=-\hbar q^2/(2m)+\hbar kq/m$. In carrying out the frequency integration care must be taken to include the contribution from the plasmon pole $\Omega_{\rm pl}$. Using the 
expression for the imaginary part of the charge-charge susceptibility near $\Omega_{\rm pl}$,
\begin{equation}
\Im m \chi_{\rm \scriptstyle C}({\bf q}, \Omega)=\pi\,v^{-1}_{\rm C}({\bf q})\left.
\Re e \chi_{0}({\bf q}, \Omega)\right|_{\Omega=\Omega_{\rm pl}}\left[\left.
\frac{\partial \Re e \chi_{0}({\bf q}, \Omega)}{\partial \Omega}\right|_{\Omega=\Omega_{\rm pl}}\right]^{-1}\delta(\Omega-\Omega_{\rm pl})\,,
\end{equation}
we find that the real part of the CH term is given by 
\begin{eqnarray}\label{conclusion_CH}
&&\left.\Re e\Sigma_{\rm \scriptstyle CH}({\bf k},\omega)\right|_{\omega=\xi_{\bf k}/\hbar}=\nonumber\\
&-&\int_{0}^{q_{\rm c}} \frac{qdq}{2\pi}\,\frac{v_{\rm C}({\bf q})\left.
\Re e \chi_{0}({\bf q}, \Omega)\right|_{\Omega=\Omega_{\rm pl}}}{\sqrt{[\Omega_{\rm pl}
+\hbar q^2/(2m)]^2-\hbar^2 k^2q^2/m^2}}\,\left[\left.
\frac{\partial \Re e \chi_{0}({\bf q}, \Omega)}{\partial \Omega}\right|_{\Omega=\Omega_{\rm pl}}\right]^{-1}\Theta(\Omega_{\rm pl}-\Omega_{\rm min}(k))\nonumber\\
&-&\int_0^{+\infty} \frac{qdq}{2\pi}\,{\mathcal P}
\int_{{\rm max}[0,\Omega_{\rm min}(k),\Omega_{\rm low}]}^{\Omega_{\rm up}}\,\frac{d \Omega}{\pi}\,
\frac{v^2_{\rm C}({\bf q})\Im m \chi_{\rm \scriptstyle C}({\bf q}, \Omega)+
3v^2_{\rm S}({\bf q})\,\Im m \chi_{\rm \scriptstyle S}({\bf q}, \Omega)}{\sqrt{[\Omega+\hbar q^2/(2m)]^2-\hbar^2 k^2q^2/m^2}}\,.
\end{eqnarray}
Here $q_{\rm c}$ marks the onset of Landau damping and $\Omega_{{\rm up}({\rm low})}=\hbar q^2/(2m)\pm \hbar qk_F/m$ are the upper and lower edges of the electron-hole continuum. 

The range of the momentum integration deserves special attention. In the first term in Eq.~(\ref{conclusion_CH}), 
due to the step function the range of $q$-integration is determined by the intersections between $\Omega_{\rm min}(k)$ and $\Omega_{\rm pl}$ (see Fig.~14). In Sect.~\ref{qp_energy} we have introduced the $r_s$-dependent wave number $k_p$: this is the wave number $k$ at which $\Omega_{\rm min}(k)$ is tangent to $\Omega_{\rm pl}$. There are two cases: (i) for $k<k_p$ there are no intersections and thus the range of $q$-integration goes from $0$ to $q_c$; and (ii) for $k\geq k_p$ there can be either one ($q_1$) or two intersections $(q_{1,2})$, so that the range of integration is $[0,q_1]$ or $[0,q_1]\cup [q_2,q_c]$, respectively. It is the 
crossover from condition (i) to (ii) that leads to the plasmon dip in 
$\left.\Re e\Sigma_{\rm \scriptstyle CH}({\bf k},\omega)\right|_{\omega=\xi_{\bf k}/\hbar}$. 

In the second term in Eq.~(\ref{conclusion_CH}) 
the range of $q$-integration runs up to $q=+\infty$ and this gives rise to the logarithmic 
divergence mentioned in Ref.~\onlinecite{footnote}. As already discussed in Sect.~\ref{quasiparticle_self_energy}, what matters are self-energy differences, which are free of singularities. Numerically we deal only with 
the finite quantity $\left.\Re e\Sigma_{\rm \scriptstyle CH}({\bf k},\omega)\right|_{\omega=\xi_{\bf k}/\hbar}-\Sigma_{\rm \scriptstyle CH}(k_F,0)$.

\section*{Appendix B. ``Line+residue" decomposition}
In this Appendix we discuss a mathematically equivalent decomposition 
of the QP self-energy, first introduced by Quinn and Ferrell~\cite{quinn}, which has been often employed
in the literature~\cite{mahan}. This amounts to writing
\begin{equation}\label{lr}
\Sigma_{\rm \scriptstyle ret}({\bf k},\omega)=\Sigma_{\rm X}({\bf k})+
\Sigma_{\rm \scriptstyle line}({\bf k},\omega)+\Sigma_{\rm \scriptstyle res}({\bf k},\omega)\,.
\end{equation}
Here the first term is the Hartree-Fock self-energy~\cite{stern_1973}
\begin{equation}\label{zero_T_HF}
\Sigma_{\rm X}({\bf k})=
\left\{
\begin{array}{l}
-2e^2 k_F\,{\rm E}(\bar{k}^2)/\pi\,\hspace{5 cm}\mbox{$(\bar{k}\leq 1)$}\\
-2e^2 k_F\,\bar{k}\,[{\rm E}(1/\bar{k}^2)-(1-1/\bar{k}^2)\,
{\rm K}(1/\bar{k}^2)]/\pi\,\hspace{0.15 cm}\mbox{$(\bar{k}>1)$}
\end{array}
\right.
\end{equation}
where $\bar{k}=k/k_F$ and ${\rm K}(x)$, ${\rm E}(x)$ are complete elliptical integrals of the first and second kind, respectively. The second term in Eq.~(\ref{lr}), which is purely real, is given by 
\begin{equation}\label{line}
\Sigma_{\rm \scriptstyle line}({\bf k},\omega)=-\int \frac{d^2 {\bf q}}{(2\pi)^2}v_{\bf q}\int_{-\infty}^{\infty} 
\frac{d\Omega}{2\pi}\left[\frac{1}{\varepsilon({\bf q},i \Omega)}-1\right]\,\frac{1}{\omega+i \Omega-\xi_{{\bf k}+{\bf q}}/\hbar}\,.
\end{equation}
Finally, the third term is the so-called ``residue" contribution,
\begin{equation}\label{residue}
\Sigma_{\rm \scriptstyle res}({\bf k},\omega)=\int \frac{d^2 {\bf q}}{(2\pi)^2} v_{\bf q} 
\left[\frac{1}{\varepsilon({\bf q},\omega-\xi_{{\bf k}+{\bf q}}/\hbar)}-1\right]\,\left[\Theta(\omega-\xi_{{\bf k}+{\bf q}}/\hbar)-\Theta(-\xi_{{\bf k}+{\bf q}}/\hbar)\right]\,.
\end{equation}
Within this decomposition it is the ``line" contribution which needs to be regularized for a ultraviolet divergence. 

As a check of our numerical results obtained by means of the SX-CH decompositions, we have recalculated the QP self-energy, effective mass, and renormalization constant by this alternative route. 
This turned out to require a substantially harder numerical effort. For completeness we summarize in Figs.~15 and~16 
our results for the ``line" and ``residue" terms.

\newpage

\begin{figure}
\begin{center}
\includegraphics[scale=0.6]{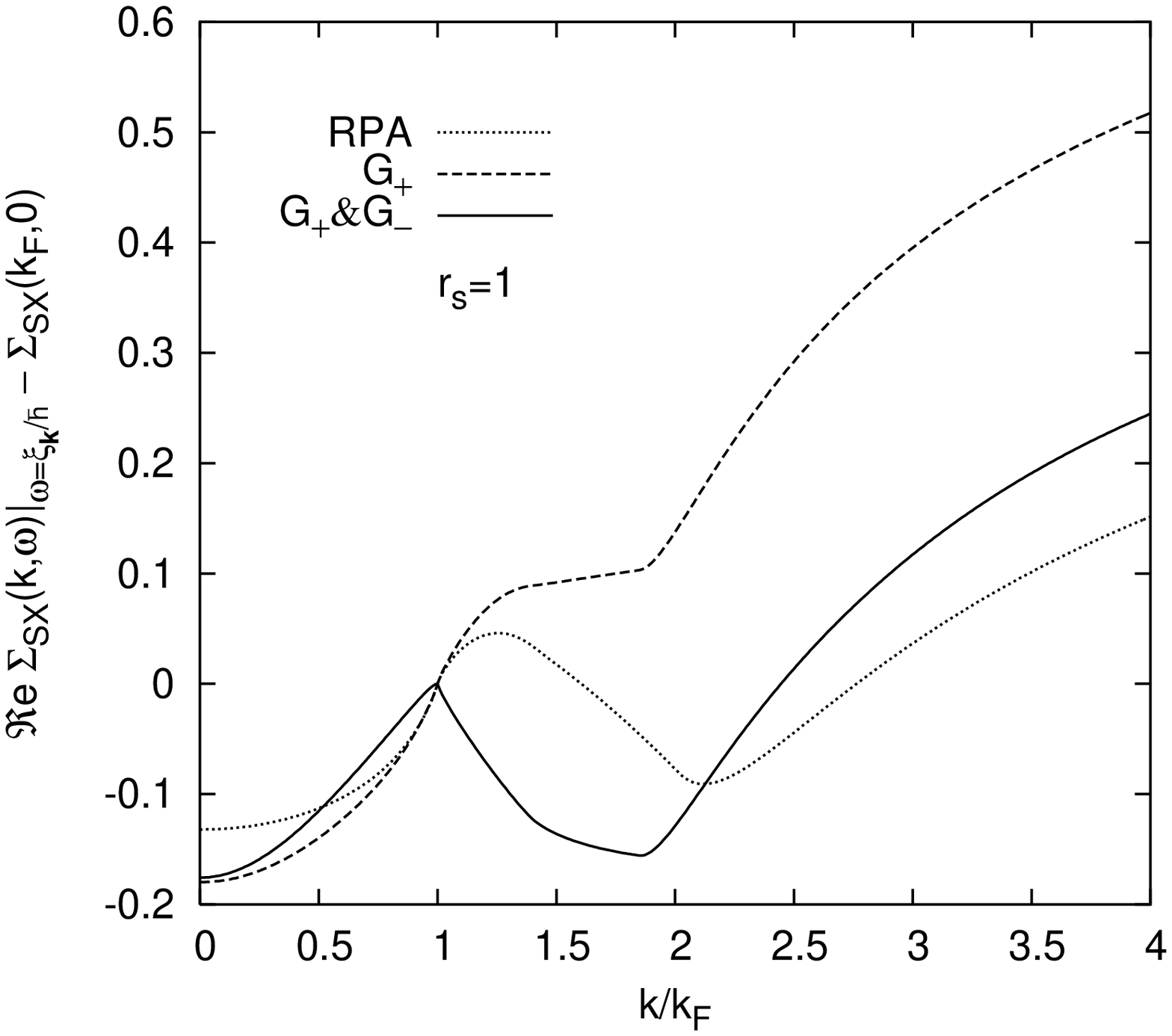}
\includegraphics[scale=0.6]{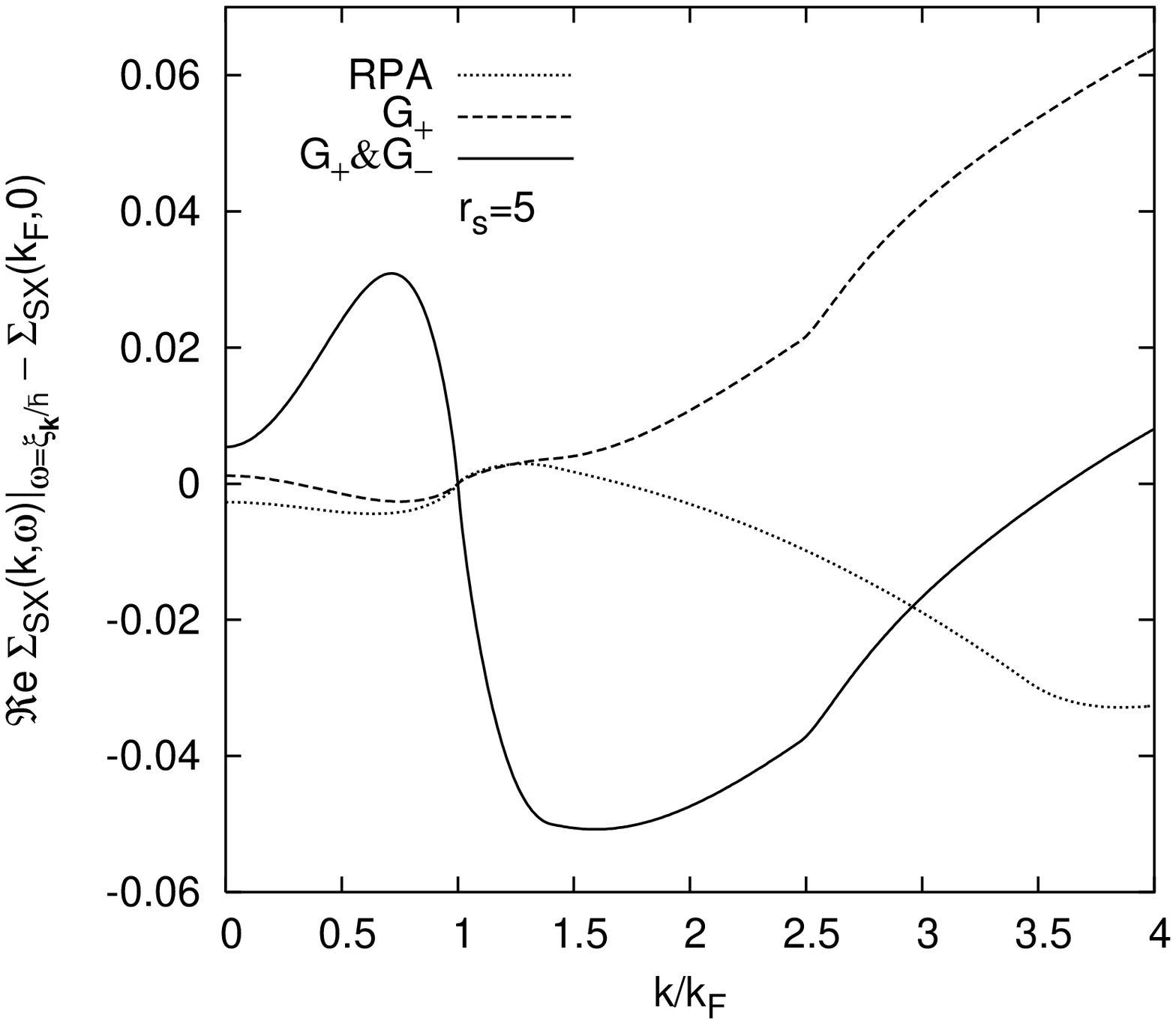}
\caption{The real part of the SX contribution to the retarded self-energy (in units of Ryd) evaluated at $\omega=\xi_{\bf k}/\hbar$, as a function of $k/k_F$ for $r_s=1$ (top panel) and $5$ (bottom panel).}
\end{center}
\label{fig1}
\end{figure}

\begin{figure}
\begin{center}
\includegraphics[scale=0.6]{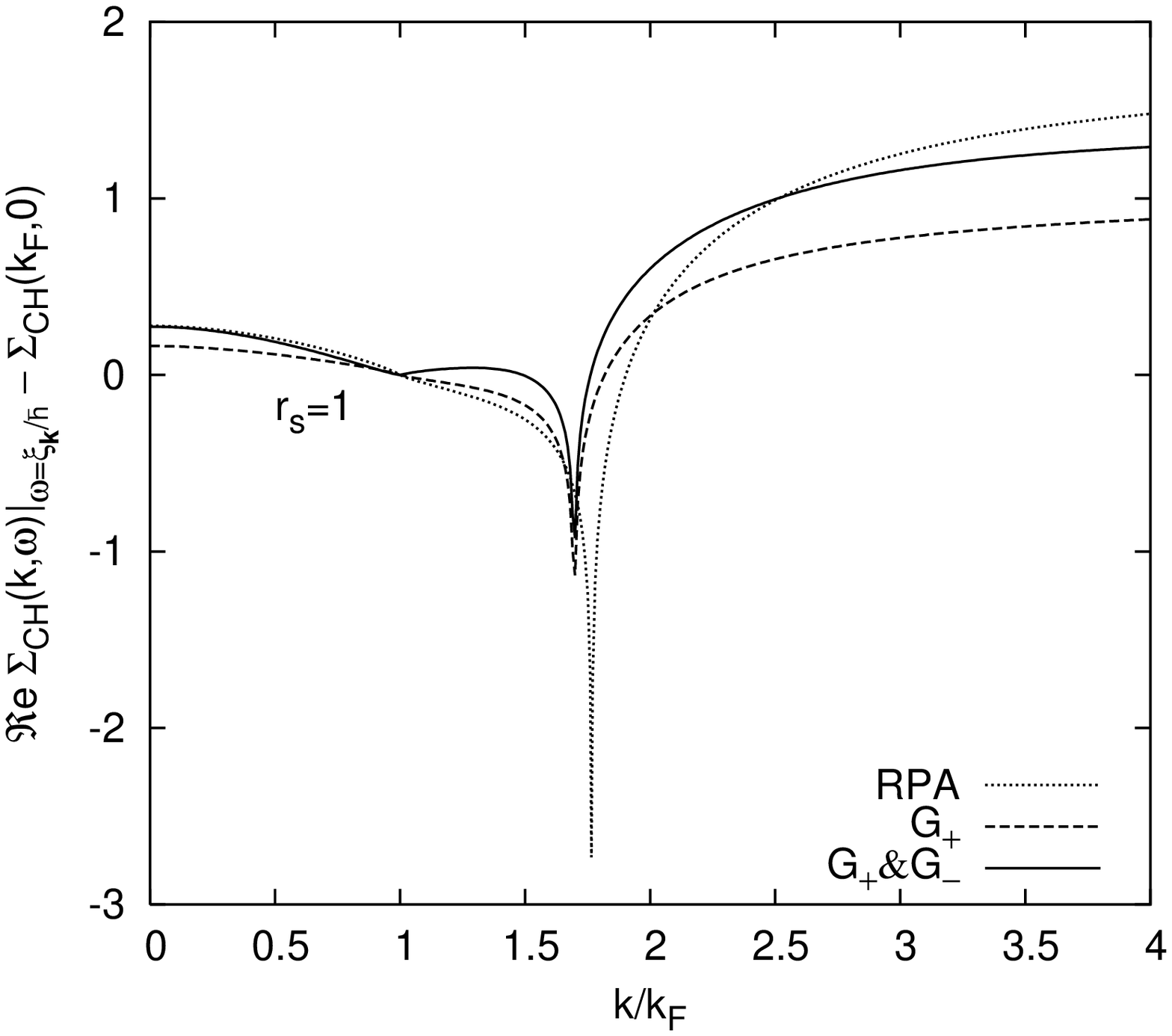}
\includegraphics[scale=0.6]{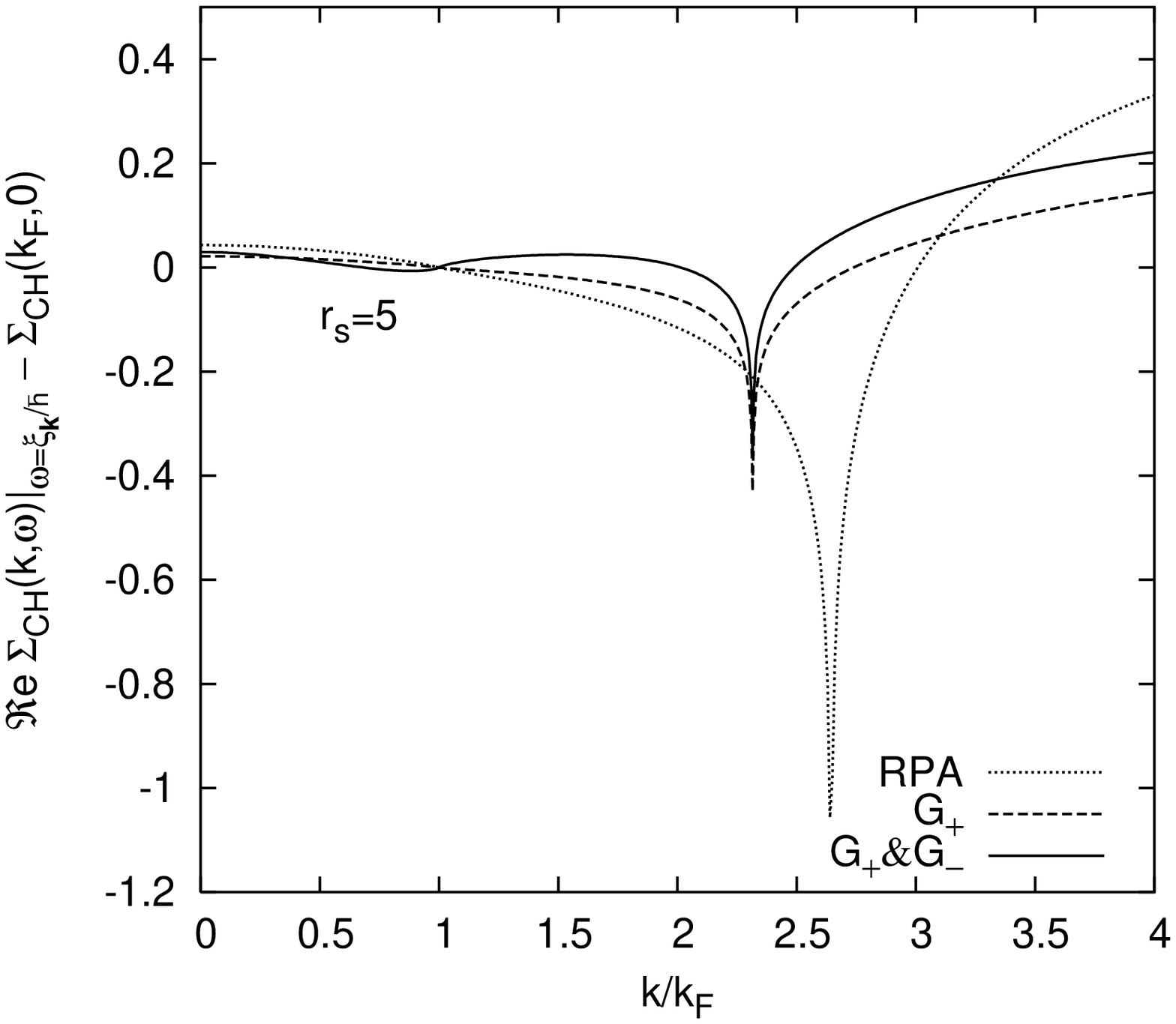}
\caption{The real part of the CH contribution to the retarded self-energy (in units of Ryd) evaluated at $\omega=\xi_{\bf k}/\hbar$, as a function of $k/k_F$ for $r_s=1$ (top panel) and $5$ (bottom panel).}
\end{center}
\label{fig2}
\end{figure}

\begin{figure}
\begin{center}
\includegraphics[scale=0.6]{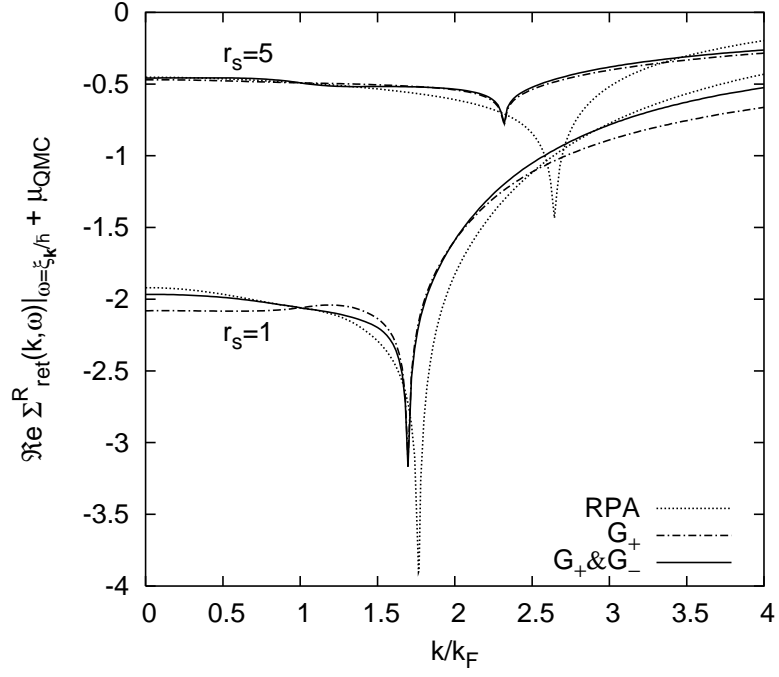}
\caption{The real part of the retarded self-energy (in units of Ryd) evaluated at 
$\omega=\xi_{\bf k}/\hbar$, as a function of $k/k_F$ for $r_s=1$ and $5$. $\mu_{\rm QMC}$ is the chemical potential from the QMC ground-state energy.}
\end{center}
\label{fig3}
\end{figure}

\begin{figure}
\begin{center}
\includegraphics[scale=0.5]{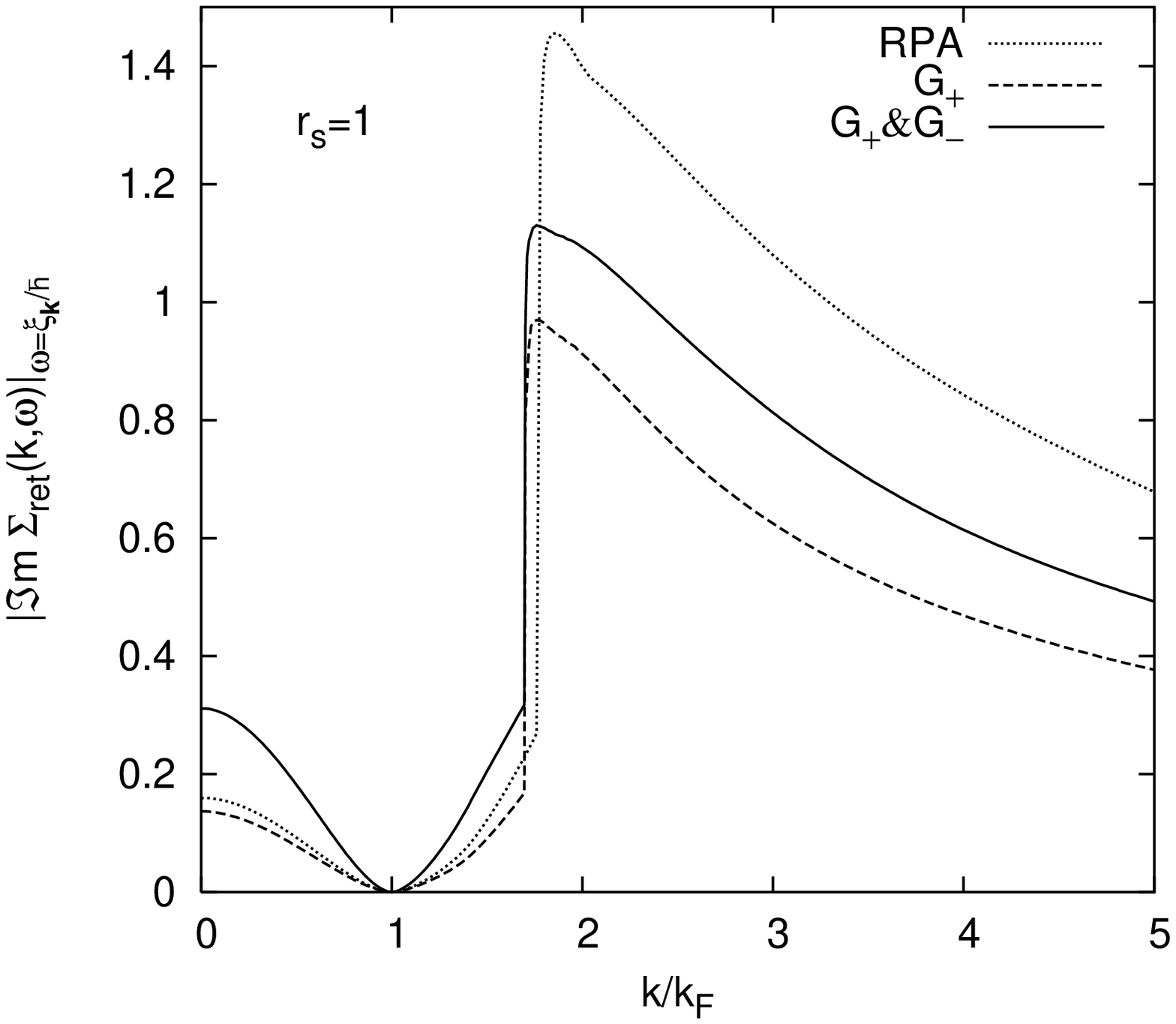}
\includegraphics[scale=0.5]{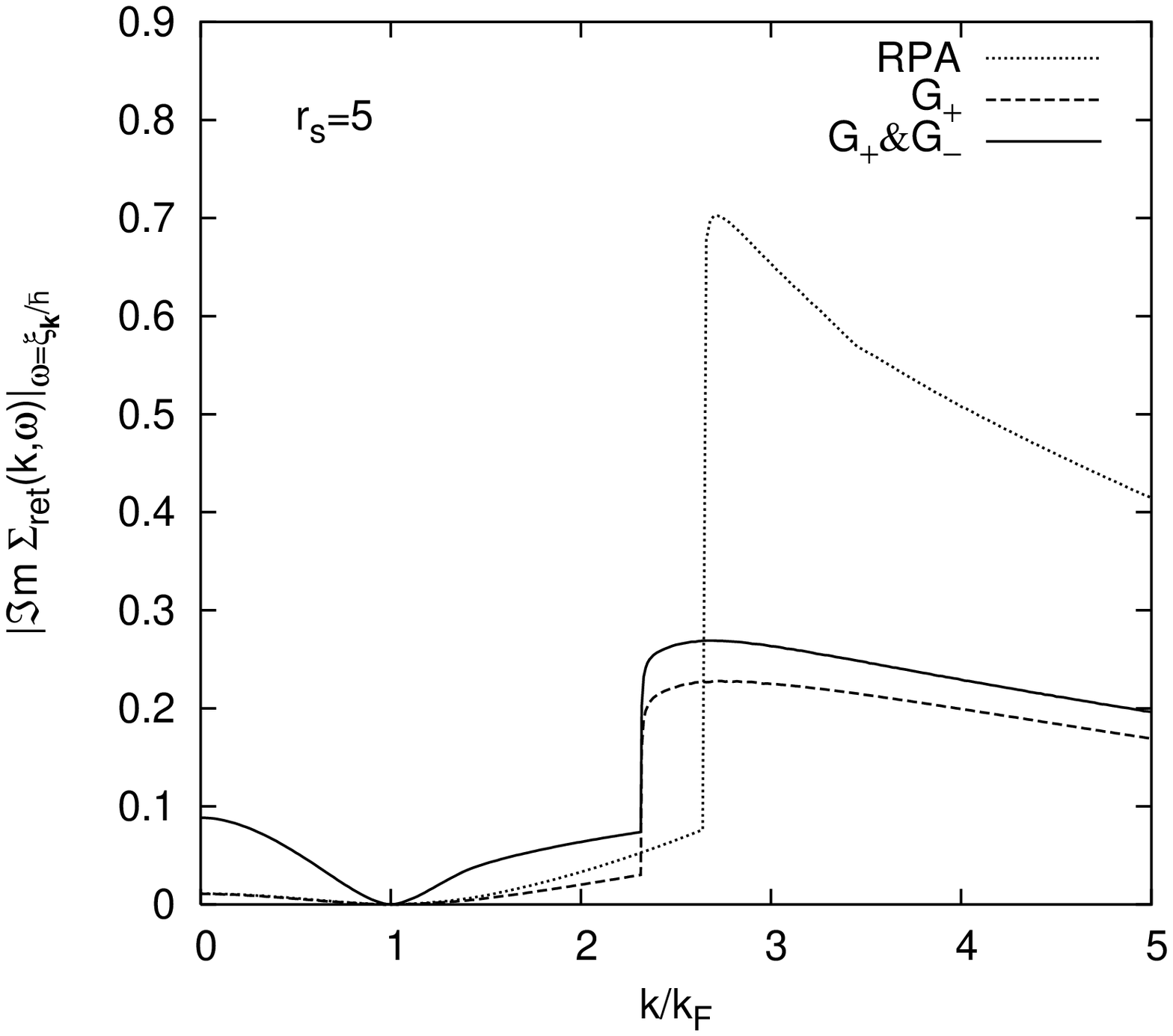}
\caption{The absolute value of the imaginary part of the retarded self-energy (in units of Ryd) evaluated at $\omega=\xi_{\bf k}/\hbar$, as a function of $k/k_F$ for $r_s=1$ (top panel) and $5$ (bottom panel).}
\end{center}
\label{fig4}
\end{figure}

\begin{figure}
\begin{center}
\includegraphics[scale=0.6]{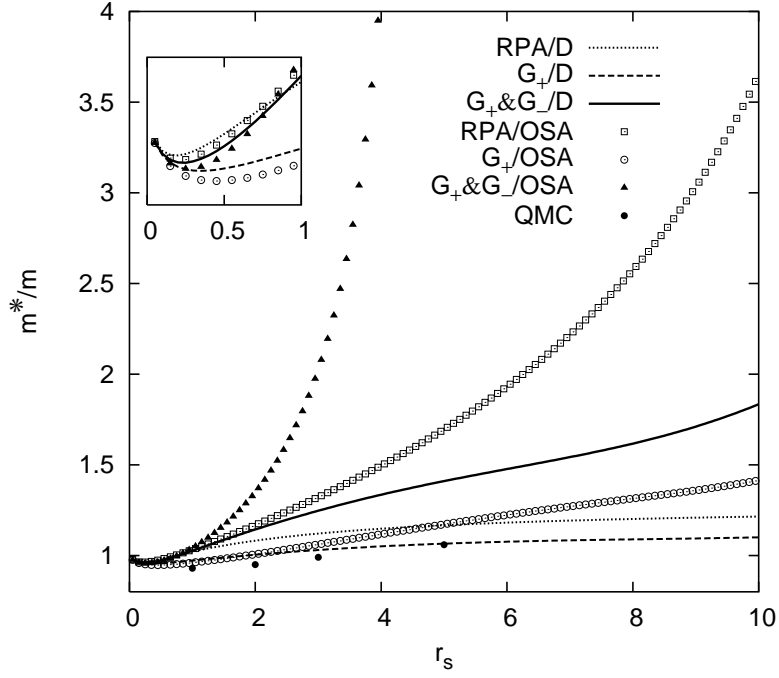}
\caption{Effective mass enhancement as a function of $r_s$ for $0 \leq r_s \leq 10$. The inset shows 
an enlargement of the results for $r_s \leq 1$. The lines show the results from Eq.~(\ref{mass_dyson}), 
while the symbols (except for the dots) are from Eq.~(\ref{mass_osa}). The QMC data (dots) are from Ref.~\onlinecite{kwon_1994}.}
\end{center}
\label{fig5}
\end{figure}

\begin{figure}
\begin{center}
\includegraphics[scale=0.6]{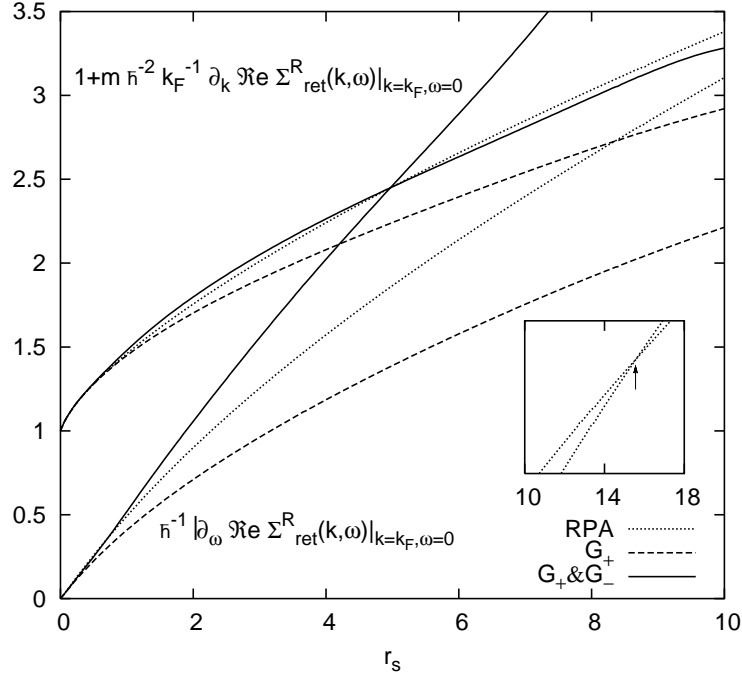}
\caption{Illustrating the divergence of the effective mass within the OSA. The 
three curves starting from unity at $r_s=0$ refer to the quantity 
$1+(m/\hbar^{2} k_F)\partial_k \Re e \Sigma^{\rm R}_{\rm \scriptstyle ret}(k_F,0)$, and 
the other three curves to $\hbar^{-1}|\partial_{\omega} \Re e \Sigma^{\rm R}_{\rm \scriptstyle ret}(k_F,0)|$. 
The intersection of two lines with the same line-style in the two sets of curves corresponds to a zero 
in the denominator of Eq.~(\ref{mass_osa}) and thus to a divergence in $m^*_{\rm OSA}$. 
The inset shows this divergence occurring within the RPA at $r_s\simeq 15.5$.}
\end{center}
\label{fig6}
\end{figure}

\begin{figure}
\begin{center}
\includegraphics[scale=0.6]{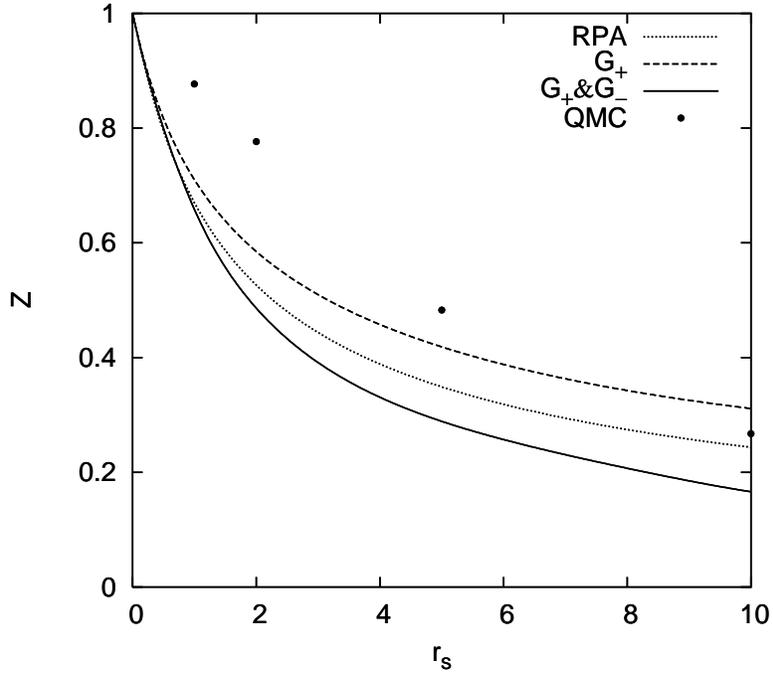}
\caption{Renormalization constant $Z$ as a function of $r_s$ for $0 \leq r_s \leq 10$. The QMC data have been taken from Ref.~\onlinecite{conti}.}
\end{center}
\label{fig7}
\end{figure}

\begin{figure}
\begin{center}
\includegraphics[scale=0.6]{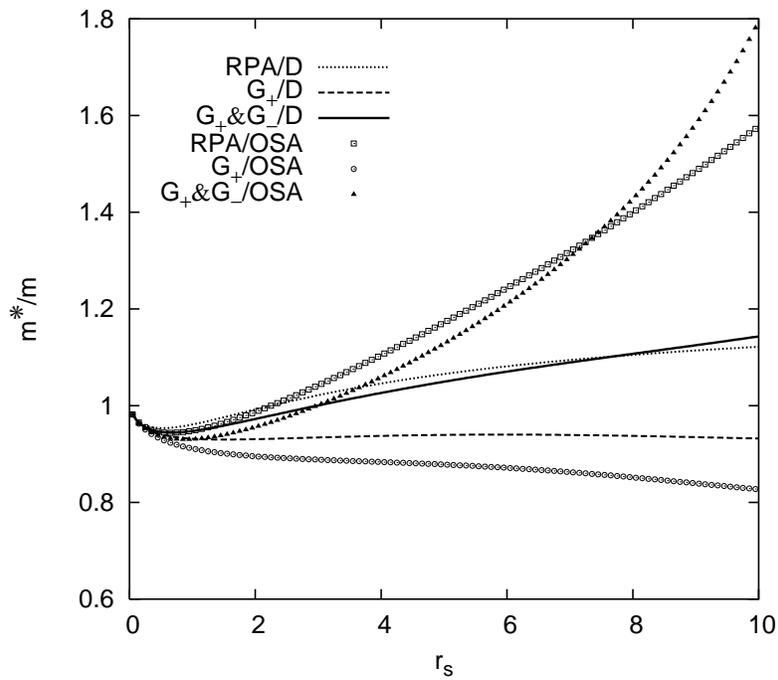}
\caption{Effective mass enhancement for a quasi-$2D$ EG confined in a GaAs/AlGaAs triangular quantum well of the type used in Refs.~\onlinecite{zhu_2003} and~\onlinecite{private_zhu}. The notation is as in Fig.~10.}
\end{center}
\label{fig8}
\end{figure}

\begin{figure}
\begin{center}
\includegraphics[scale=0.6]{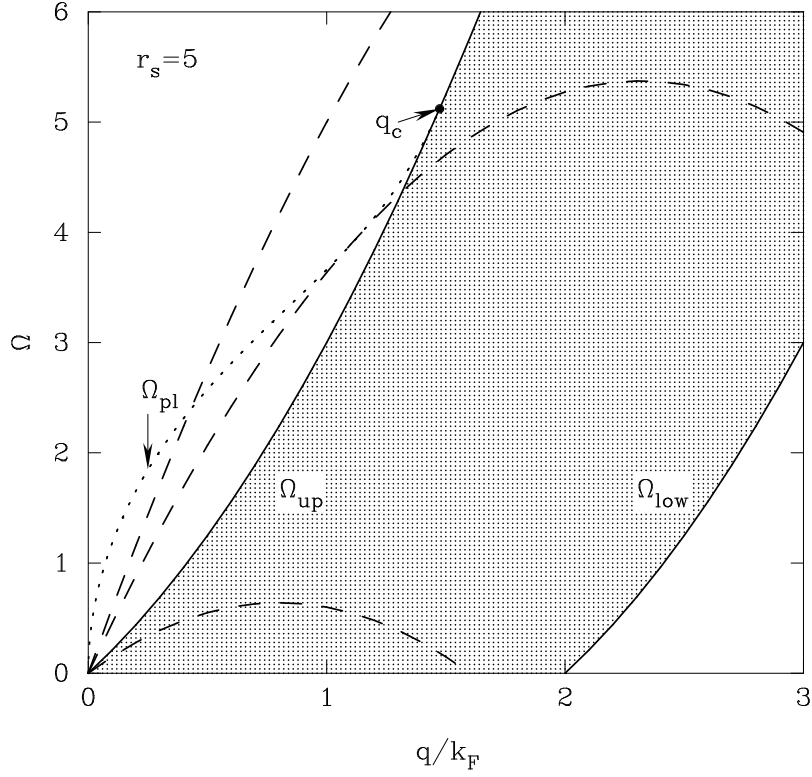}
\caption{Illustrating the integration range relevant to Eq.~(\ref{conclusion_CH}), with $\Omega$ in units of $\varepsilon_F/\hbar$. The shaded area represents the electron-hole continuum. $\Omega_{\rm min}(q)$ 
is shown for three values of $k$  (dashed curves): from top to bottom, $k=3.0k_F,k_p$, and $0.8k_F$ 
with $k_p\simeq 2.32k_F$. The plasmon dispersion relation $\Omega_{\rm pl}(q)$ (dotted curve) is also shown.}
\end{center}
\label{fig9}
\end{figure}

\begin{figure}
\begin{center}
\includegraphics[scale=0.7]{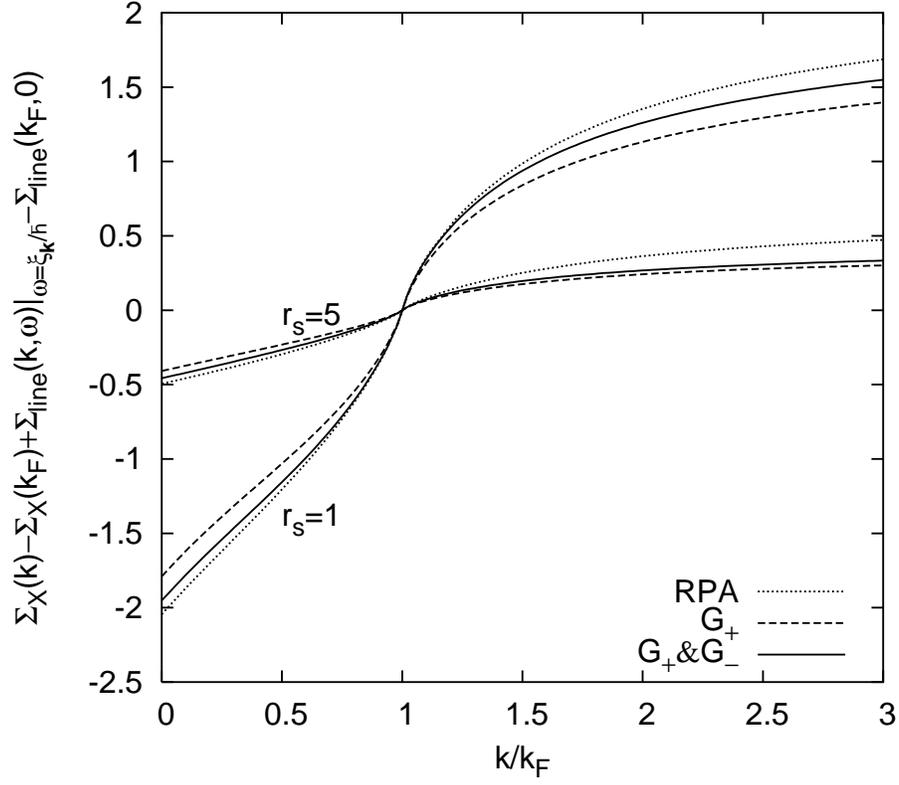}
\caption{The exchange plus regularized ``line" contribution to the retarded self-energy (in units of Ryd) evaluated at $\omega=\xi_{\bf k}/\hbar$, as a function of $k/k_F$ for $r_s=1$ and $5$.}
\end{center}
\label{fig10}
\end{figure}

\begin{figure}
\begin{center}
\includegraphics[scale=0.5]{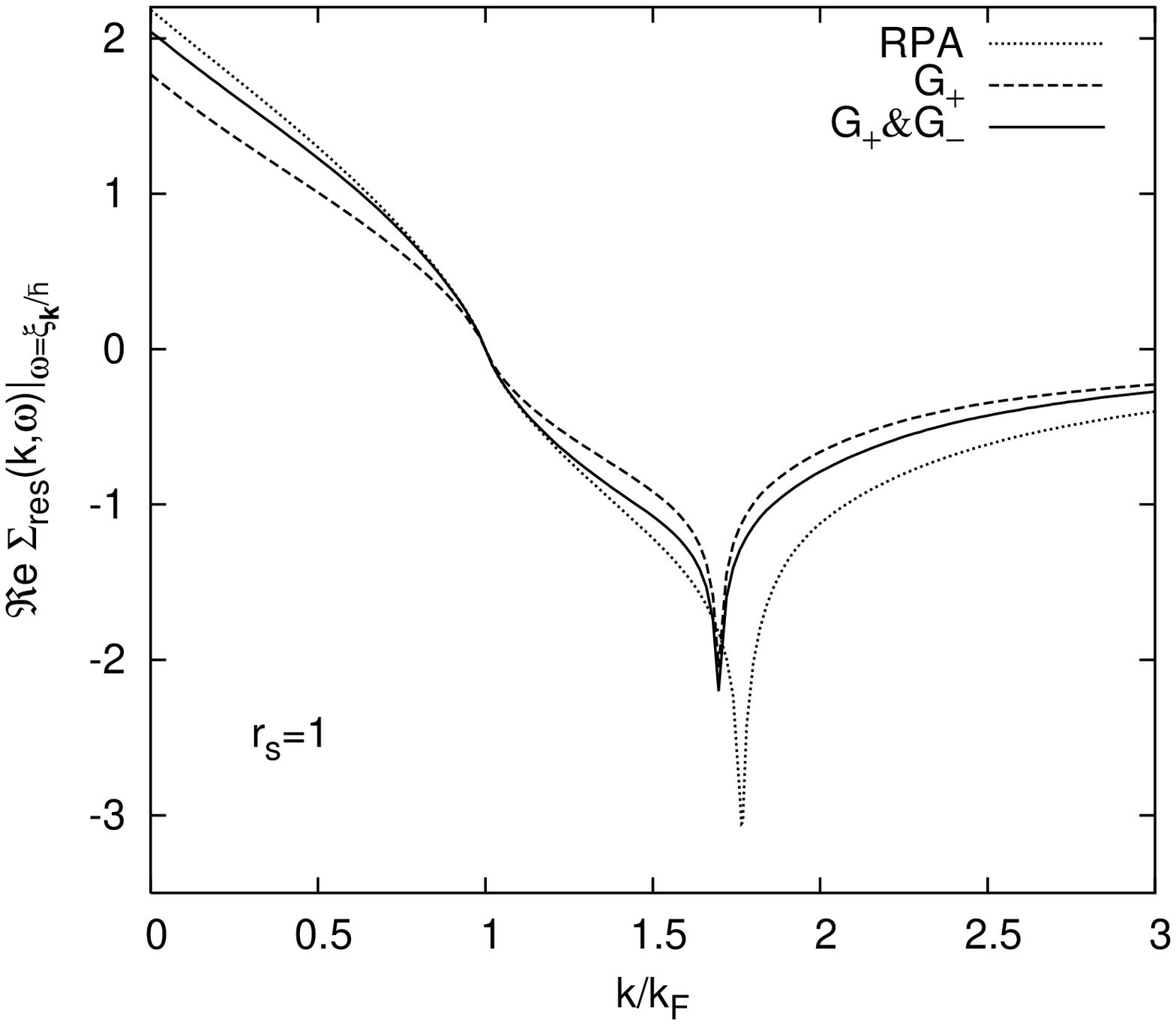}
\includegraphics[scale=0.5]{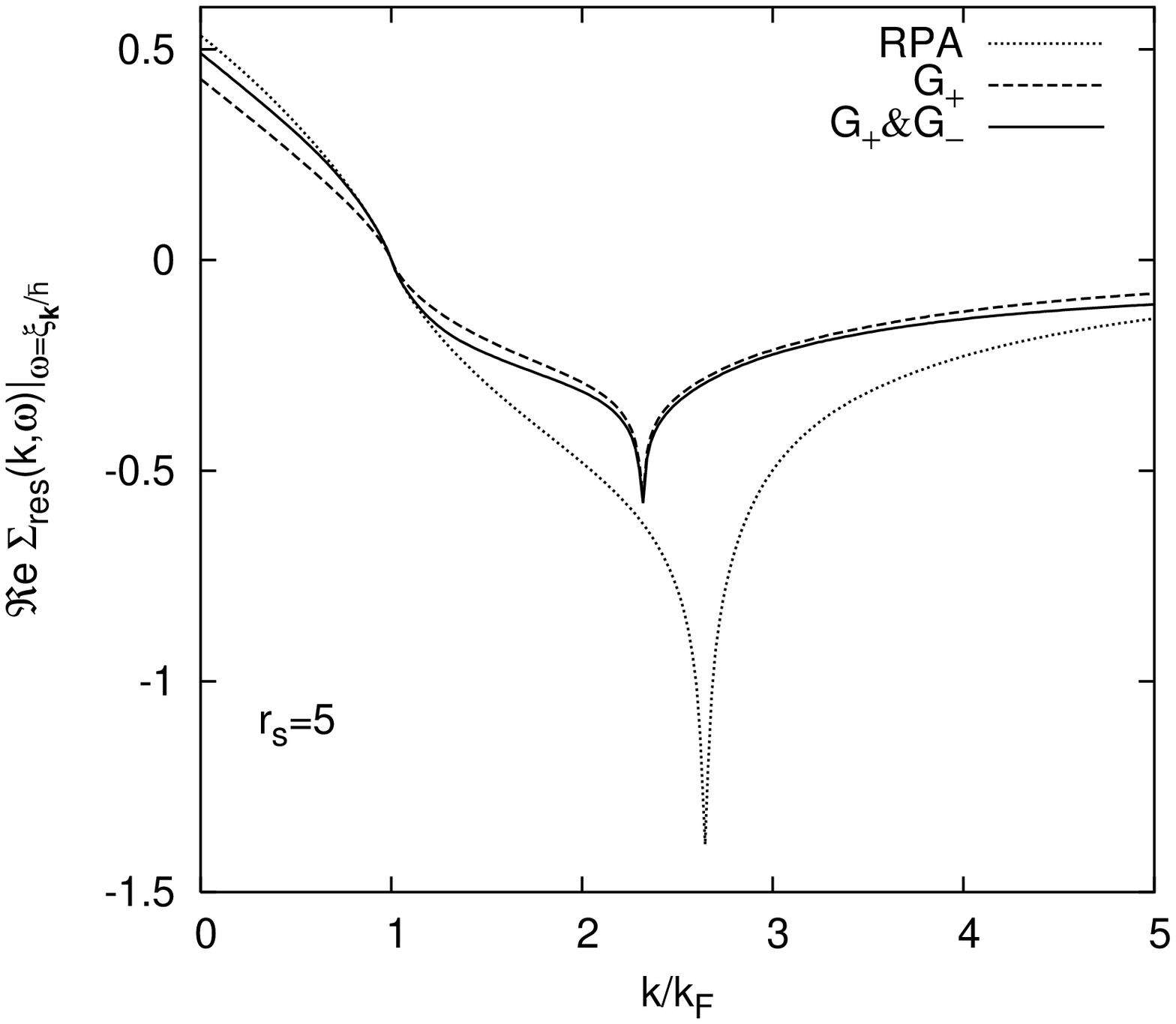}
\caption{The real part of the ``residue" contribution to the retarded self-energy (in units of Ryd) evaluated at 
$\omega=\xi_{\bf k}/\hbar$, as a function of $k/k_F$ for $r_s=1$ (top panel) and $5$ (bottom panel).}
\end{center}
\label{fig11}
\end{figure}

\newpage

\begingroup
\squeezetable
\begin{table}
\caption{QP effective mass and renormalization constant of a paramagnetic $2D$ EG. The arrows refer to the divergence in the OSA as explained in the main text.}
\begin{ruledtabular}
\begin{tabular}{llllllll}
$r_s$ &Various calculations   &$m^*_{\rm D}/m$ & $m^*_{\rm OSA}/m$ & $Z$\\ \hline
1			&RPA	 		 						&1.022 &1.033 &0.670\\
			&$G_+$						  	&0.972 &0.961 &0.710\\
			&$G_+ \& G_-$		 	    &1.026 &1.040 &0.658\\ \hline
2			&RPA		  						&1.082 &1.168 &0.526\\
			&$G_+$						  	&1.004 &1.007 &0.585\\
			&$G_+ \& G_-$		 	    &1.144 &1.349 &0.486\\ \hline
3			&RPA		  						&1.121 &1.322 &0.444\\
			&$G_+$						    &1.030 &1.061 &0.510\\
			&$G_+ \& G_-$	 	  		&1.247 &2.026 &0.391\\ \hline
5			&RPA		  						&1.167 &1.696 &0.349\\
			&$G_+$							  &1.066 &1.172 &0.419\\
			&$G_+ \& G_-$		 	  	&1.410 &$\nearrow$ &0.289\\ \hline
10    &RPA		  						&1.215 &3.650 &0.244\\
			&$G_+$							  &1.100 &1.415 &0.311\\
			&$G_+ \& G_-$		 	  	&1.834 &$\nearrow$ &0.166\\ 
\end{tabular}
\end{ruledtabular}
\end{table}
\end{document}